\documentclass[12pt]{article}
\usepackage{latexsym,epsfig,amssymb,amsmath,subfigure,color,pictex,fullpage,amsfonts,ulem}
\usepackage{amssymb}
\usepackage{authblk}

\usepackage{lineno}

\usepackage{latexsym}
\usepackage{epsfig}
\usepackage{amsmath}
\usepackage{subfigure}
\usepackage{pictex}
\usepackage{fullpage}
\usepackage{amsfonts}
\usepackage{float}
\usepackage[colorlinks=true]{hyperref}
\restylefloat{table}

\usepackage{xcolor}

\newcommand{\ma}[1]{\textcolor{magenta}{#1}}

\newcommand{\ie}{{\it i.e.}\ }
\newcommand{\eg}{{\it e.g.}\ }
\newcommand{\etal}{{\it et~al.}\ }
\newcommand{\e}[1]{\ensuremath{\times 10^{#1}}}
\newcommand{\ds}{\displaystyle}

\newcommand{\eps}{\epsilon}

\newcommand{\tp}{t_0^+}
\newcommand{\De}{D_{\rm{e}}}
\newcommand{\be}{\begin{eqnarray}}
\newcommand{\ee}{\end{eqnarray}}
\newcommand{\bes}{\begin{eqnarray*}}
\newcommand{\ees}{\end{eqnarray*}}

\newcommand{\Fms}{{N}_{-}}

\newcommand{\phie}{\Phi}

\newcommand{\je}{j}

\newcommand{\jn}{j_n}

\newcommand{\ja}{j_{a}}

\newcommand{\caa}{c_{a}}

\newcommand{\etaa}{\eta_a}
\newcommand{\phia}{\Phi_a}
\newcommand{\Da}{D_{a}}
\newcommand{\Ra}{R_{a}}
\newcommand{\ueqa}{U_{eq,a}}
\newcommand{\cam}{c_a^{\rm max}}
\newcommand{\jc}{j_{c}}

\newcommand{\cc}{c_{c}}

\newcommand{\etac}{\eta_c}
\newcommand{\phic}{\Phi_{c}}
\newcommand{\Dc}{D_{c}}
\newcommand{\Rc}{R_{c}}
\newcommand{\ueqc}{U_{eq,c}}
\newcommand{\ccm}{c_c^{\rm max}}

\usepackage[margin=1in]{geometry}

\title{DandeLiion v1: {An extremely fast} solver for the Newman model of lithium-ion battery (dis)charge}

\author[1,2,5]{Ivan Korotkin}
\author[2,3,6]{Smita Sahu}
\author[2,4,7]{Simon~O'Kane}
\author[1,2,8]{Giles Richardson}
\author[2,3,9]{Jamie M. Foster}
\affil[1]{Mathematical Sciences, University of Southampton, University Rd., SO17 1BJ, UK}
\affil[2]{The Faraday Institution, Quad One, Becquerel Avenue, Harwell Campus, Didcot, OX11 0RA, UK}
\affil[3]{School of Mathematics and Physics, University of Portsmouth, Lion Terrace, PO1 3HF, UK}
\affil[4]{Department of Mechanical Engineering, Imperial College London, Exhibition Road, SW7~2AZ, UK}
\affil[5]{\tt i.korotkin@soton.ac.uk}
\affil[6]{\tt smita.sahu@port.ac.uk}
\affil[7]{\tt s.okane@imperial.ac.uk}
\affil[8]{\tt g.richardson@soton.ac.uk}
\affil[9]{\tt jamie.michael.foster@gmail.com}

\begin{document}

\maketitle

\begin{abstract}
DandeLiion (available at {\tt dandeliion.com}) is a robust and extremely fast solver for the Doyle Fuller Newman (DFN) model, the standard electrochemical model for (dis)charge of a planar lithium-ion cell. DandeLiion conserves lithium, uses a second order spatial discretisation method (enabling accurate computations using relatively coarse discretisations) and is many times faster than its competitors. The code can be used `in the cloud' and does not require installation before use. The difference in compute time between DandeLiion and its commercial counterparts is roughly a factor of 100 for the moderately-sized test case of the discharge of a single cell. Its linear scaling property means that the disparity in performance is even more pronounced for bigger systems, making it particularly suitable for applications involving multiple coupled cells. The model is characterised by a number of phenomenological parameters and functions, which may either be provided by the user or chosen from DandeLiion's library. This library contains data for the most commonly used electrolyte (LiPF$_6$) and a number of common active material chemistries including graphite, lithium iron phosphate (LFP), nickel cobalt aluminium (NCA), and a variant of nickel cobalt manganese (NMC).

\end{abstract}


%% main text

\section{Introduction}
\label{intro}

Lithium-ion batteries (LIBs) provide rechargeable energy storage at an unrivalled energy and power density, with a high cell voltage, and a slow loss of charge when not in use \cite{Blo17}. These characteristics have lead to their widespread use {in consumer} electronics, and their increasing dominance in electric vehicle (EV) applications and off grid storage. Driven largely by the incumbent legislation to ban the combustion engine across large parts of the world before 2040, it has been predicted that the demand for LIBs will balloon from 45~GWh/year (in 2015) {to 390~GWh/year} in 2030 \cite{Zub18}. Thus, the need to {improve and optimise LIB technology is especially timely and, in particular, the development of underpinning modelling capabilities promises to significantly accelerate this process}. Particularly in {the case of EV applications significant challenges remain. These are associated with the demanding requirements made of vehicle batteries, including long service life,  rigorous safety standards and good performance under aggressive charge/discharge regimes \cite{Vet05,Wan12}}. 

\begin{figure} \centering
\includegraphics[width=0.95\textwidth]{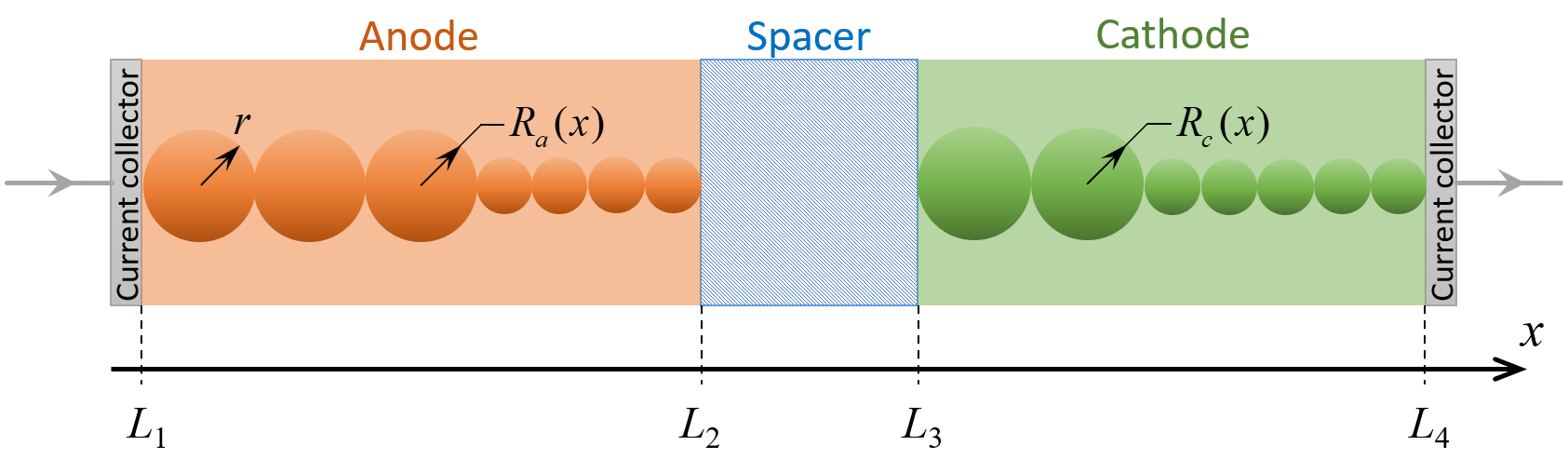}
\caption{Schematic of a planar LIB cell. The macroscopic and microscopic coordinates, $x$ and $r$, are indicated, along with the particle radii and positions between the different cell components.}
\label{cell_figure}
\end{figure}

{A single LIB cell consists of two porous electrodes (an anode and a cathode) separated by a porous spacer (see figure \ref{cell_figure}) and sandwiched between two current collectors. The cell is bathed in a liquid lithium electrolyte that acts to transport charge, and lithium, between the two electrodes. Each electrode is comprised of an agglomeration of electrode particles formed from active materials into which lithium ions can intercalate. For modelling purposes electrode particles are often assumed to be spherical. Under discharge conditions Li$^+$ ions, which have a greater chemical energy in the anode material than the cathode material, deintercalate from the anode particles, migrate through the electrolyte and across the porous separator to the cathode where they intercalate into the cathode particles. The transport of charge, from anode to the cathode, that results from this migration of the positively charged Li-ions gives rise to a potential difference, between the two electrodes, that can be used to drive a current through an external circuit.}

{The electrochemistry and electrical behaviour of a LIB cell is typically modelled by the Doyle Fuller Newman (DFN) model \cite{doyle96,Doy93,Ful94}. This describes the charge transport and Li-ion migration within the cell. In particular it couples nonlinear diffusion models for Li-ion transport within the electrode particles to a semi-phenomenological model for the electrolyte, which is able to accurately capture ion transport and electrical conduction, via a Butler-Volmer model \cite{Ful94,NewmanBook} that quantifies the rate of (de-)intercalation of Li-ion from the surfaces of the electrode particles. More details of this model, and its relationship to the underlying physics and chemistry of the device, can be found in \cite{NewmanBook,New75,RicRev}.}

{Although the DFN model is to some extent the gold standard in engineering simulations of LIBs it is nevertheless prohibitively computationally expensive in many applications. In particular, composite cells such as pouch and cylindrical cells have heterogeneous temperature distributions and therefore required a DFN solution to be carried out at each point in space and coupled to a three-dimensional heat transport equation. This results in a five-dimensional problem that is extremely compuationally challenging. Various approaches are adopted to reduce the complexity of this problem including equivalent circuit modelling of the cell and more recently systematic (single particle) asymptotic reductions of the DFN model which reduce the dimension of the DFN model by one, see \cite{spm_paper,marquis19}. Rather than simplify the model, our approach here is to tailor highly-efficient numerical methods to the DFN model and thereby reduce the computational time to a level that is acceptable for the type of computations that we might wish to perform.}

%Lithium transport within individual electrode particles is modelled by a nonlinear diffusion equation and electronic conduction between particles within the electrode by Ohm's law. Lithium ion transport and ionic conduction within the electrolyte is assumed to obey the standard electrolyte model (see \eg \cite{NewmanBook,New75,RicRev}) which consists of a nonlinear diffusion equation for the Li-ion concentration and a concentration-dependent variant of Ohm's law. \ma{The reaction that transports lithium ions across the surface of the electrode particles, between the active material and the electrolyte}, is, as usual, described by the Butler-Volmer equation \cite{Ful94,NewmanBook}.

{The numerical software presented in this work (DandeLiion v1) is designed to solve the DFN model and is motivated by the pressing need for fast, and powerful, numerical code that is capable of solving computationally expensive problems in battery design, such as the simulation of the thermally coupled electrochemical behaviour of composite cells (\eg pouch cells and jelly-roll cells), battery modules, and (even) entire battery packs. It also has the potential to significantly enhance other computationally expensive tasks such as the optimisation of cell design and estimation of parameters from experimental data.}

The rest of this work is devoted to the description of the numerical procedure, adopted in DandeLiion, for the solution of the DFN model \cite{doyle96,Doy93,Ful94} of charge transport in a single LIB cell. 
%Their future development will be significantly accelerated by improvements in modelling capability. 

{\subsection{Software performance and operation} The DandeLiion solver is based upon a method of lines approach to the solution of the system of mixed parabolic-elliptic partial differential equations (PDEs) that comprise the DFN model. In this approach the DFN PDEs are first discretized in space to yield a large system of coupled time-evolving ODEs and algebraic equations. This system of differential algebraic equations (DAEs) is efficiently solved using an in-house solver, written by the authors and based upon Backward Differentiation Formulae and adaptive time stepping \cite{arieh,suli}.}

{The DandeLiion solver} has been validated against (i) in-house code implemented in MATLAB \cite{matlab}, (ii) experiments and simulations described in the work of Ecker \etal \cite{Ecker2015a,Ecker2015b}, and alternative implementations of the DFN model in both (iii) the Battery Library in Dymola \cite{dymola}, a proprietary code, and (iv) PyBaMM, an open source project \cite{pybamm}. For the cross-verification with Dymola, the code was parametrised using the same model and battery properties as described in \cite{dymola_paper}. {In \cite{spm_paper} it has been compared to (v) an approximate, simplified, reduced-order battery cell model, showing very good agreement between the two different approaches,} even for relatively high discharge rates up to around 12C. An example comparison between PyBaMM, experiment and DandeLiion is shown below in \S\ref{examples} and further work \cite{spm_paper,Alana} also verify DandeLiion against other experiments and simulations.
%\red{(I CAN IMAGINE THAT WE WILL BE ASKED TO BACK UP SOME OF THESE ASSERTIONS)} \bl{\it{We have a comparison with Ecker, reduced-order model and PyBaMM in \cite{spm_paper}. We even have some comparison with PyBaMM and Ferran's experiments in this paper below (in Illustrative examples section).}}
%\red{(CAN WE  BACK UP THESE ASSERTIONS)} \bl{\it{It depends on many factors, how much faster. We may trigger even more questions and discussions from refs if we start to give numbers here. I think, we should be careful and avoid saying that other solvers are rubbish. Let's focus on DandeLiion and say that it is very fast because of linear scaling, 2nd order scheme, C++ implementation, etc. And we can give an example that the full DFN model, one single discharge can be solved during a fraction of second (see figure 2, where 7000 DAEs correspond to 0.1 sec.)}

{The DandeLiion solver}  works {very much faster} than {our previous MATLAB implementation of a DFN solver} (which is based on {\tt ode15s}), the implementation of the model in the Dymola Battery Library \cite{dymola} and than the open source implementation PyBaMM \cite{pybamm}. To quantify this, our MATLAB implementation, which is comparable in speed with Dymola, was outperformed by DandeLiion by a factor of around 100 in terms of reduced computational time for a single cell discharge. The disparity in performance is significantly greater for more larger problems such as pouch and {cylindrical cells where the linear scaling properties of DandeLiion become even more pronounced.} {Furthermore, in contrast to Dymola and PyBaMM, DandeLiion uses a second order spatial discretization and therefore requires many fewer space points to achieve the same accuracy as these other solvers, which are only first order accurate in space.} A full discharge cycle of a battery at a moderate (1C) discharge rate that involves solution of a system of approximately 7000 coupled nonlinear DAEs takes {less than a second of simulation time for DandeLiion on a standard desktop computer. For comparison, solution of the same problem takes around one minute, or even more in MATLAB, using the same hardware, and this gap in code performance becomes more pronounced for bigger systems.} The number of DAEs that {the DandeLiion solver can handle on a desktop computer with 16~Gb of RAM (available in most standard desktops) is enough to solve approximately $2\e{7}$ DAEs and can be increased beyond $10^8$ depending on the machine's RAM. Furthermore code performance is not hampered when the number of DAEs increases, {and the simulation time scales {\it linearly} with the number of DAEs being solved (see Figure \ref{scaling_figure}). This is in contrast to most other DFN codes whose simulation times scale {\it quadratically} with system size. Such codes are therefore prohibitively computationally expensive when used for large computations.} Importantly, this opens {the possibility of simulating multidimensional systems, 3D composite cells, such as a pouch cell which is made by stacking a large number of individual cells (typically around 50) on top of each other. Heat generation within such cells can lead to significant heterogeneities in the temperature distribution, which in turn leads to heterogeneities in the electrochemical properties (which are highly sensitive to temperature) of individual cells. Since the DFN model for a single isothermal cell is two-dimensional in space (one micro dimension measuring distance from the centre of an electrode particle and one macroscopic cell dimension measuring distance across the cell) the thermally coupled model that needs to be solved for a composite cell is five dimensional (3 macroscopic pack dimensions, 1 macroscopic cell and 1 micro dimension). Such problems are extremely computationally challenging and require fast and efficient solvers, such as DandeLiion. Other computationally intensive applications for which efficient code is highly desirable include parameter estimation and cell optimisation {routines, both of which require that multiple simulations,} using different sets of parameters, are performed on a single cell.}

\begin{figure} \centering
\includegraphics[width=0.6\textwidth]{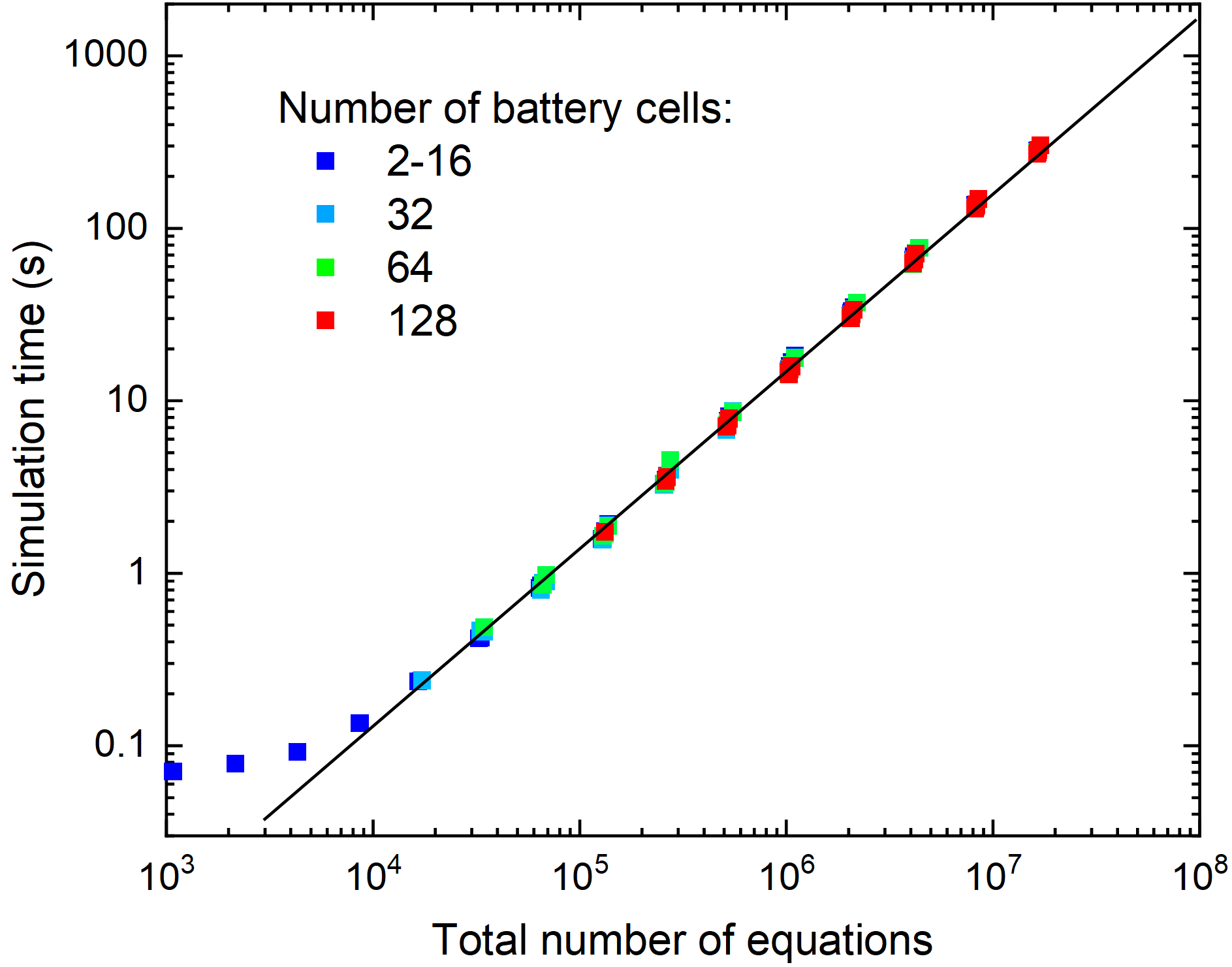}
%\caption{Simulation time per equation vs the total number of DAEs}
\caption{
{Simulation time in seconds vs the total number of differential-algebraic equations (DAEs) in the system. In this test, a stack of up to 128 battery cells (DFN models) has been simulated at 1C discharge rate from fully pre-charged state until fully discharged using different computational grids. Each point in the plot represents a separate simulation with one particular grid size. The density of the spatial grid varies from relatively coarse, 10 nodes in the electrolyte and 50 in each electrode particle, up to relatively fine, 1280 nodes in the electrolyte and 800 nodes in each particle. Each battery cell in a stack was parametrised according to \cite{Ecker2015a,Ecker2015b}. All cells are connected via potentiostatic boundary conditions (parallel battery connection). The test was performed on a desktop with {\tt Intel Core i7-8700} CPU and 16~Gb of RAM.}}
\label{scaling_figure}
\end{figure}
% explain non-linear behaviour at the beginning, expected time at 1e8, swap file after 2e7 equations.

{DandeLiion does not require installation and is available to be used for computations `in the cloud' at {\tt dandeliion.com}. The website hosts comprehensive documentation as well as a series of tutorial videos aimed at educating new users on how to use the tool, which cover a range of topics, including how to simulate a full discharge cycle, modification of the model parameters and functions, how to simulate drive cycles, creating and adding a user-defined electrode chemistries, and setting up graded electrodes {(electrodes with different particle sizes)}. Several pre-defined examples are available, and these are intended to serve as {templates which can be adapted for specific customised simulations thereby lowering the barrier to entry for new users}.

\section{The DFN model and software implementation}
\label{description}
%Describe the software in as much as is necessary to establish a vocabulary needed to explain its impact. 

DandeLiion is a framework for simulating {LIB cell charge and discharge. It solves  a 1+1D (pseudo 2D)} DFN porous electrode model that was established in \cite{Doy93,doyle96,Ful94} {and is reviewed in detail in \cite{RicRev}}. The current version allows the user to choose from a library of preprogrammed parameterisations for the electrolyte and electrode chemistries or, alternatively, to specify their own parametrisations. The model parameters may be changed by the user by editing a simple web form. Complete web forms submit a simulation to the queue and computations are carried out on our dedicated server free of charge. A concise view of the results of finished computations can be viewed live in the web browser and we also provide an option for users to download the raw output of simulations for in-depth analysis using the software of their choice.

The core DandeLiion code is written in C++ which is partially responsible for its fast performance. It is based on a 2nd order spatial discretization of the system of PDEs which comprise the DFN model. A finite element discretisation is employed in the macroscopic dimension, $x$ {(which measures distance across the cell)}, as described in detail below and in the microscopic dimension, $r$ {(which measures distance from the centre of each electrode particle)}, we use a control volume method described in \cite{Zang13}. These methods are both 2nd order accurate and, crucially, are conservative. The latter point means that when they are applied in tandem to the macroscopic and microscopic lithium conservation equations, {that form part of the DFN model,} they ensure global lithium conservation throughout the device. This property is particularly important when multiple (dis)charge cycles are performed and ensures that the battery's capacity is retained.

{The finite element and control volume methods are applied to the DFN model equations and in doing so spatial derivatives are removed. The spatially-discretised DFN model can be written as a large coupled system of time-dependent DAEs, i.e. $\mathbf{M}\dot{\mathbf{u}} = \mathbf{f}(\mathbf{u})$, where ${\mathbf u}$ is a vector containing the time-dependent values of the model variables at the computational grid points, the mass matrix is denoted by $\mathbf{M}$, and the ``right-hand side function'' is denoted by $\mathbf{f}(\mathbf{u})$, see equation (\ref{30}).} DAE systems are typically more problematic to solve than a system comprised solely of coupled ODEs \cite{wanbook}, but there are commercial solvers aimed at solving such systems, such as MATLAB's ode15s \cite{cel14}. Solving the DAEs that result from the spatial discretisation of the DFN model is by far the most computationally expensive part of the solution procedure and for this reason we have developed {a specialised in-house} DAE solver, as part of the DandeLiion code, which is based on implicit variable-order (2 to 6) backward differentiation formulae \cite{arieh,suli} and an optimised Newton root-finding method.

%\red{\sout{It employs a simple but powerful C++ DAE solver {\tt dae-cpp} that uses only one external dependency: the Intel Math Kernel Library \cite{mkl}, a fast and very well optimised (especially for Intel-based systems) linear algebra library available from the Intel website.}} \bl{Maybe we want to get rid of this text}

\subsection{The DFN Model}
{In what follows we lay out the full cell 1+1d DFN model {\cite{doyle96,Doy93,Ful94} that is solved by DandeLiion}; for a more detailed description of the physics and chemistry underlying this model the reader is referred to {Newman's book \cite{NewmanBook} and} the review article \cite{RicRev}.  The version of the 1+1d DFN model considered here describes a one dimensional cell lying between $x=L_1$ and $x=L_4$ (see figure \ref{cell_figure}), consisting of}
\bes
\begin{array}{lc}
\mbox{an anode in} & L_1<x<L_2,\\
\mbox{a separator in} & L_2<x<L_3,\\
\mbox{and a cathode in} & L_3<x<L_4.
\end{array}
\ees
The model comprises one-dimensional macroscopic equations  posed  across the width of the cell  $L_1<x<L_4$. These describe electrical conduction in the solid matrices of the anode and cathode and lithium ion transport and conduction in the electrolyte that fills the pores of the electrode matrix. They couple to one-dimensional spherically symmetric microscopic lithium transport equations posed posed in representative spherical electrode particles, {which occupy the regions $0 \leq r <R_a(x)$ in the anode and $0 \leq r <R_c(x)$ in the cathode. Here $R_a(x)$ and $R_c(x)$, {which are allowed to vary in space to allow for the possibility of particle grading}, give the radii of the anode and cathode particles, respectively,} as a function of $x$. The full cell DFN model is formulated below in equations \eqref{fc1}-\eqref{fc51}; the associated model variables are listed, and described, in Table \ref{table1}, and the model parameters and functions are catalogued in Table \ref{table2}.

\begin{table}[ht]
\centering
\begin{tabular}{|c|c|c|}
\hline \textbf{Variable} & \textbf{Description} & \textbf{Units} \\
\hline $x$ & Distance across cell & m \\
\hline $t$ & Time & s \\
\hline $r$ & Distance from centre of electrode particle & m \\
\hline $c$ & Ion concentration in electrolyte & mol\,m$^{-3}$ \\
\hline $\Fms$ & Average flux of negative counterions in electrolyte & mol\,m$^{-2}$s$^{-1}$ \\
\hline $\Phi$ & Electric potential w.r.t. lithium electrode in electrolyte & V \\
\hline $j$ & Average current density in electrolyte & A\,m$^{-2}$ \\
\hline $j_n$ & Current density on surface electrode particles & A\,m$^{-2}$ \\
             & flowing from electrode particle to electrolyte & \\
\hline $j_a$ & Average current density in anode & A\,m$^{-2}$ \\
\hline $j_c$ & Average current density in cathode & A\,m$^{-2}$ \\
\hline $\Phi_a$ & Electric potential in anode & V \\
\hline $\Phi_c$ & Electric potential in cathode & V \\
\hline $\caa$ & Lithium-ion concentration in anode particles & mol\,m$^{-3}$ \\
\hline $\cc$ & Lithium-ion concentration in cathode particles & mol\,m$^{-3}$ \\
\hline $\eta_a$ & Overpotential between electrolyte and anode particles & V \\
\hline $\eta_c$ & Overpotential between electrolyte and cathode particles & V \\
\hline $V(t)$ & Potential difference across device & V \\
\hline
\end{tabular}
\caption{Description of the variables used in the formulation of the full cell DFN model}
\label{table1}
\end{table}

\begin{table}[ht]
\centering
\begin{tabular}{|c|c|c|}
\hline \textbf{Param./} & \textbf{Description} & \textbf{Units} \\
\textbf{Ftn.}  &  &  \\
\hline $T$ & Absolute temperature & K \\
\hline ${\cal B}(x)$ & Permeability factor in electrode matrix & dim'less \\
\hline $\epsilon_l(x)$ & Volume fraction of electrolyte in electrode matrix & dim'less \\
\hline $b(x)$ & Brunauer-Emmett-Teller (BET) surface area & m$^{-1}$ \\
\hline $\De(c)$ & Ionic diffusivity of electrolyte: function of concn. & m$^2$s$^{-1}$ \\
\hline $\tp$ & Transference number & dim'less \\
\hline $\kappa(c)$ & Electrolyte conductivity as function of concentration & A\,m$^{-1}$V$^{-1}$ \\
\hline $\sigma_a(x)$ & Anode conductivity as function of concentration & A\,m$^{-1}$V$^{-1}$ \\
\hline $\sigma_c(x)$ & Cathode conductivity as function of concentration & A\,m$^{-1}$V$^{-1}$ \\
\hline $\Ra(x)$ & Radius of anode particles as function of position & m \\
\hline $\Rc(x)$ & Radius of cathode particles as function of position & m \\
\hline $\cam$ & Max. lithium concentration in anode particles & mol\,m$^{-3}$ \\
\hline $\ccm$ & Max. lithium concentration in cathode particles & mol\,m$^{-3}$ \\
\hline $k_a$ & Butler-Volmer constant in anode & mol$^{-1/2}$m$^{5/2}$s$^{-1}$ \\
\hline $k_c$ & Butler-Volmer constant in cathode & mol$^{-1/2}$m$^{5/2}$s$^{-1}$ \\
\hline $\ueqa(\caa)$ & Open-circuit voltage: function of Li$^+$ concn. in anode & V \\
\hline $\ueqc(\cc)$ & Open-circuit voltage: function of Li$^+$ concn. in cathode & V \\
\hline $\Da(\caa)$ & Li$^+$ diffusivity anode: function of Li$^+$ concn. & m$^2$ s$^{-1}$ \\
\hline $\Dc(\cc)$ & Li$^+$ diffusivity cathode: function of Li$^+$ concn. & m$^2$ s$^{-1}$ \\
\hline $I(t)$ & Current flow into cell & A \\
\hline ${\cal R}_{cont}$ & Total contact resistance & V\,A$^{-1}$ \\
\hline $A$ & Electrode cross-sectional area & m$^2$ \\
\hline $c_0$ & Initial ionic concentration in electrolyte & mol\,m$^{-3}$ \\
\hline $c_{a,0}$ & Initial ionic concentration in anode & mol\,m$^{-3}$ \\
\hline $c_{c,0}$ & Initial ionic concentration in cathode & mol\,m$^{-3}$ \\
\hline
\end{tabular}
\caption{User specified functions and parameters for full cell DFN model}
\label{table2}
\end{table}

\paragraph{The Macroscopic equations} 
\be \label{fc1}
\epsilon_l(x)\frac{\partial c}{\partial t}+\frac{\partial \Fms}{\partial x}=0,\quad 
\Fms=-\mathcal{B}(x)\De(c)\frac{\partial c}{\partial x}-(1-\tp)\frac{\je}{F} \quad \mbox{in} \quad L_1 < x < L_4. ~~~~~~\\
\label{fc2} \frac{\partial \je}{\partial x}= b(x) \jn, \qquad 
\je=-\mathcal{B}(x)\kappa(c)\left(\frac{\partial \phie}{\partial x}-\frac{2RT}{F}\frac{1-\tp}{c}\frac{\partial c}{\partial x}\right)\quad \mbox{in} \quad L_1 < x< L_4, ~~~~~~\\
\label{fc3}
\frac{\partial \ja}{\partial x}=-b(x) \jn,\quad 
\ja=-{\sigma_a}\frac{\partial \phia}{\partial x} \quad \mbox{in} \quad L_1< x<L_2, ~~~~~~~~~~~~~~~~~~\\
\frac{\partial \jc}{\partial x}=-b(x) \jn,\quad 
\jc=-{\sigma_c}\frac{\partial \phic}{\partial x} \quad \mbox{in} \quad L_3< x<L_4, ~~~~~~~~~~~~~~~~~~\\
\label{fc4}
\jn=\left\{\begin{array}{ccc}
\ds 2 F k_a c^{1/2} \left( \caa \rvert _{r=\Ra(x)} \right)^{1/2} \left( \cam -\caa \rvert_{r=\Ra(x)}\right)^{1/2} \sinh\left(\frac{F \etaa}{2RT}\right) &  \mbox{in} & L_1 \leq x < L_2, \\
0 & \mbox{in} & L_2 < x < L_3, \\
\ds 2 F k_c c^{1/2} \left( \cc \rvert _{r=\Rc(x)} \right)^{1/2} \left( \ccm -\cc \rvert_{r=\Rc(x)}\right)^{1/2} \sinh\left(\frac{F\etac}{2RT}\right) &  \mbox{in} & L_3 \leq x < L_4, \end{array} \right.\\
\label{fc7} \etaa=\phia-\phie-\ueqa(\caa|_{r=\Ra(x)}), \qquad \etac=\phic-\phie-\ueqc(\cc|_{r=\Rc(x)}). ~~~~~~~~~
\ee
Varibles in the anode and cathode are distinguished by their subscripts: we use the variables $\phia$, $\ja$, $\caa$, $\eta_a$ in the anode ($L_1<x<L_2$) and $\phic$, $\jc$, $\cc$, $\eta_c$ in the cathode ($L_3<x<L_4$). Furthermore the electrode particles in the anode and cathode have different electrical properties and so are characterised by different equilibrium potential functions, $\ueqa(\caa)$ in the anode and $\ueqc(\cc)$ in the cathode.

\paragraph{Macroscopic boundary and interface conditions}
Here the macroscopic boundary and interface conditions on the model are
\be
\label{fc21} \ja|_{x=L_1} &=& \frac{I(t)}{A}, \quad \Fms|_{x=L_1} = 0, \quad j|_{x=L_1} = 0,\\
\label{fc22} \ja|_{x=L_2} &=& 0, \\
\label{fc23} \jc|_{x=L_3} &=& 0, \\
\label{fc24} \jc|_{x=L_4} &=& \frac{I(t)}{A}, \quad \Fms|_{x=L_4} = 0, \quad j|_{x=L_4} = 0.
\ee
representing galvanostatic discharge at a current $I(t)$ which flows into the anode current collector on $x=L_1$ through the anode particles and out through the cathode current collector on $x=L_4$ through the cathode particles (figure \ref{cell_figure}). No electronic current passes through the electronically insulating separator.

\paragraph{Microscopic equations and boundary conditions}
The microscopic equations and boundary conditions on the model are given by
\be
\label{fc30}
\left. \begin{array}{l} \ds \frac{\partial \caa}{\partial t}=\frac{1}{r^{2}}\frac{\partial }{\partial r}\left( r^{2} \Da (\caa)\frac{\partial \caa}{\partial r}\right) \quad \mbox{in} \quad 0< r<\Ra(x)\\
\ds \caa \,\, \mbox{bounded} \,\, \mbox{on} \,\, r=0, \qquad -\Da(\caa) \frac{\partial \caa}{\partial r}\bigg\rvert_{r= \Ra(x)}=\frac{\jn}{F} 
\end{array} \right\}\ \  \mbox{in} \ \  L_1< x<L_2, \\
\label{fc31} \left. \begin{array}{l} \ds \frac{\partial \cc}{\partial t}=\frac{1}{r^{2}}\frac{\partial }{\partial r}\left( r^{2} \Dc (\cc)\frac{\partial \cc}{\partial r}\right) \quad \mbox{in} \quad 0< r<\Rc(x)\\
\ds \cc \,\, \mbox{bounded} \,\, \mbox{on} \,\, r=0, \qquad -\Dc(\cc) \frac{\partial \cc}{\partial r}\bigg\rvert_{r= \Rc(x)}=\frac{\jn}{F} 
\end{array} \right\}\ \  \mbox{in} \ \  L_3< x<L_4,
\ee
where $D_a$ and $D_c$ are the diffusivities of Li$^+$ in the of the anode and cathode particles respectively.

\paragraph{Initial conditions}
{Constant initial conditions are provided for the ion concentration in the electrolyte
\be\label{fc32}
c|_{t=0}= c_0,
\ee
and likewise for those in the active materials in the anode and cathode
\be\label{fc33}
%\caa|_{t=0}= (U_{til}-C_{SEI}) c_{c,max} \frac{\int_{L_3}^{L_4}(1-\epsilon_l(x)) a_{ct,c} dx}{\int_{L_1}^{L_2}(1-\epsilon_l(x)) a_{ct,a} dx} , \qquad \cc|_{t=0}= (1-U_{til}) c_{c,max}.
\caa|_{t=0}= c_{a,0}, \qquad \cc|_{t=0}= c_{c,0},
\ee
where $c_0$, $c_{a,0}$, and $c_{c,0}$ are provided by the user.}
%and an explanation for these formulae can be found in \cite{Ecker2015b}.

% Combined with Table 2
%\begin{table}[ht]
%\centering
%\begin{tabular}{|c|c|c|}
%\hline \textbf{Param./} & \textbf{Description} & \textbf{Units} \\
%\textbf{Ftn.} & & \\
%\hline
%$C_{SEI}$ & Fraction of capacity lost due to SEI formation in anode & dim'less \\
%\hline $U_{til}$ & Initial degree of utilization of cell & dim'less \\
%\hline $a_{ct,a}$ & Fraction of anode matrix that is active material & dim'less \\
%\hline $a_{ct,c}$ & Fraction of cathode matrix that is active material & dim'less \\
%\hline $c_0$ & Initial ionic concentration in electrolyte & mol\,m$^{-3}$ \\
%\hline
%\end{tabular}
%\caption{Parameterisation of global lithium content}
%\label{table3}
%\end{table}

\paragraph{The full cell potential}
The results of solution to the full cell DFN model and a specified galvanostatic current $I(t)$ can be used to compute the potentials at the anode and cathode current collectors $V_a$ and $V_c$, respectively via the relations
\be \label{fc50}
V_a(t)= \phia\big\rvert_{x=L_1}, \qquad V_c(t)= \phic\big\rvert_{x=L_4}.
\ee
and hence the potential drop across the full cell (\ie the cell voltage) is given by
\be
V(t)=V_c(t)-V_a(t)-{\cal R}_{cont} I(t).   \label{fc51}
\ee
{where ${\cal R}_{cont}$ is the contact resistance of the cell.}

\subsection{Software functionalities}
\label{functionalities}
%Present the major functionalities of the software.

In the most basic case the user can specify a current draw from/supply to the cell and {the code} will solve for the internal concentration, potential and current density profiles as well as the cell voltage during discharge/charge until the device reaches a user-defined cut-off potential. The results of a simulation in such a scenario are discussed in \S\ref{examples}. However, DandeLiion can also be used in a number of more sophisticated ways and it has the capability to simulate: (i) a variety of cell chemistries, (ii) {realistic drive cycles, (iii) graded electrodes in which particle size varies across the electrode, and {(iv) GITT (Galvanostatic Intermittent Titration Technique) experiments}}. The user may refer to the `Getting Started' page on {\tt dandeliion.com} for tutorials.

\paragraph{{Materials and Electrolyte library}} Data for the electrolyte LiPF$_6$ in the form of functions for electrolyte diffusivity $\De(c)$, electrolyte conductivity $\kappa(c)$ and a value for the transference number {$\tp$} is provided in the DandeLiion's parameter library. Similarly data is provided for the electrode materials graphite (Li$_x$C$_6$), LNC ($\text{Li}_x(\text{Ni}_{0.4}\text{Co}_{0.6})\text{O}_2$), LFP ($\text{Li}_x\text{FePO}_4$), {and NMC ($\text{Li}\text{Ni}_{1-x-y}\text{Mn}_x\text{Co}_{y}\text{O}_2$)}. For each of these materials we provide the open circuit voltage $U_{eq}(c_s)$ as a function of the concentration of intercalated lithium. For graphite and LNC we also provide the lithium diffusivity $D_s(c_s)$ within the material as a function of concentration of intercalated lithium. In the case of $\text{LiFePO}_4$ we assume a constant diffusivity value {$D_s=8 \times 10^{-18}$ m$^2$s$^{-1}$} \cite{Newman_LFP}. In future releases this library will be expanded enlarging the choice of pre-defined chemistries and materials. All other parameter values and functions are taken from \cite{Ecker2015a,Ecker2015b,Ferran}.

%\red{(DO WE NEED TO GIVE REFERENCES FOR WHERE THE DATA FOR THE LIBRARY COMES FROM??)} \bl{Added a reference for LFP, other refs are given in the library itself.} \red{I am not quite sure what is meant by the `library itself'. I checked the simulation submission page and references are not give there. Can we be more specific?}

% IK: I'm afraid we don't have enough space to describe this. I referred to the tutorial web page instead.
%\paragraph{Simulating a drive cycle}
%\paragraph{Simulating layered electrodes}

\section{{Spatial discretisation of the DFN model}} %}\label{fem}
{In this section we discuss the second order spatial discretisation of the DFN model \eqref{fc1}-\eqref{fc51} that leads to the system of DAEs that are solved by DandeLiion. This is based on a control volume method for the microscopic equations \eqref{fc30}-\eqref{fc31} that has been previously given in \cite{Zang13} and a novel finite element method for the macroscopic equations \eqref{fc1}-\eqref{fc24}, which we describe in detail below.}

\subsection{{\label{fem} Finite element discretisation of the macroscopic equations}}
Here, we discuss the spatial second-order finite element discretisation for the macroscopic equations \eqref{fc1}-\eqref{fc7}, {for the electrolyte and current transport in the solid parts of the anode and cathode}. A similar method has been used for a related systems of equations describing charge transport in solar cells in \cite{Courtier18}. The microscopic diffusion equations \eqref{fc30}-\eqref{fc31} are discretised using the {conservative control volume method given in Zeng \etal \cite{Zang13} which is chosen both for its second order accuracy, which matches the rate convergence of the scheme that we use for the macroscopic equations, and because it gives direct access to the concentration on the surface of the particle without the need for extrapolation. This latter feature is particularly important because the charge transfer reaction rate, given by the Butler-Volmer equations \eqref{fc4}, depends strongly upon the surface concentration of lithium and any errors made in computing its value gives rise to large errors in the transfer current undermining the quality of the simulations.}

For a positive integer $N$, let $\{L_1=x_0<x_1...<L_4=1\}$ be a partition of $[L_1,L_4]$ into the subintervals $(x_{i-1},x_{i}), ~ 1\leq i\leq N$ with grid spacing $\Delta_{i+\frac{1}{2}}= x_{i+1}-x_{i}$. The computational grid is comprised of $N+1$ points. We apply the approach described in \cite{Johnsonbook} to derive the finite element descretisation. The idea is to approximate dependent variables as a linear combination of piecewise linear basis functions (aka `hat' or `tent' functions). For a generic dependent variable, say $w$, we write
\be\label{fem0}
w(x,t)= \sum_{i=0}^{i=N}w_i(t)\psi_i(x) \quad  \text{where} \quad \psi_i(x)= \left\{
\begin{array}{ll}
	\frac{x-x_{i-1}}{x_i-x_{i-1}}& x\in (x_{i-1},x_i) \\
	\frac{x_{i+1}-x}{x_{i+1}-x_{i}}& x\in (x_i,x_{i+1}) \\
	0 & x\notin (x_{i-1},x_{i+1})\\
\end{array} 
\right.,
\ee
in which $\psi_i(x)$ is referred to as the basis functions. {By eliminating $\Fms$, $\je $, $\ja$, and $\jc$}
%the flux of anions in the electrolyte, $\Fms$, the current density in electrolyte, $\je $, the current density in anode, $\ja$, and current 
%density in cathode, $\jc$, 
from equations \eqref{fc1}-\eqref{fc7}, we obtain {the following set of macroscopic PDEs:} 
\be\label{fem1}
\eps_l(x)\frac{\partial c}{\partial t}+\frac{\partial}{\partial x}\left(  q_1(x,c) \frac{\partial c}{\partial x} + q_2(x,c)  \frac{\partial \phie}{\partial x}  \right)=0,\\
\label{fem2}
\frac{\partial }{\partial x}\left(  q_3(x,c) \frac{\partial c}{\partial x} + q_4(x,c)  \frac{\partial \phie}{\partial x}  \right)  =b(x)\jn,\\
\label{fem3}
{\sigma_a}\frac{\partial}{\partial x}\left(\frac{\partial \phia}{\partial x}\right)=b(x)\jn,\\
\label{fem4}
{\sigma_c}\frac{\partial}{\partial x}\left(\frac{\partial \phic}{\partial x}\right)=b(x)\jn,
\ee
in which
\be\label{coeff}
\begin{split}
	q_1(x,c) &= -\mathcal{B}(x)\left(\De(c) + \kappa(c)\frac{2RT(1-\tp)^2}{F^2c}\right),\quad
	q_2(x,c)  =\mathcal{B}(x)\kappa(c)\frac{(1-\tp)}{F},\\
	q_3(x,c)  &=\mathcal{B}(x)\kappa(c)\frac{2RT(1-\tp)}{Fc},\quad \text{and}\quad
	q_4(x,c)  =-\mathcal{B}(x)\kappa(c).
\end{split}
\ee
These are to be solved subject to the boundary conditions (\ref{fc21})-(\ref{fc24}). Each of the equations (\ref{fem1})-(\ref{fem4}) have the following form
\be\label{fem5}
\gamma_1(x) \frac{\partial  w}{\partial t } =  \frac{\partial }{\partial x} \left(\gamma_2(x,w)\frac{\partial w }{\partial x} + \gamma_3(x,w)\frac{\partial \phi }{\partial x} \right) + S(x,v,w).
\ee
and, in the interests of brevity, we shall now discuss how the finite element method is applied to (\ref{fem5}) {rather than discussing each of equations} (\ref{fem1})-(\ref{fem4}) individually. 
%In order to get equation \eqref{fem1}-\eqref{fem4},  If we substitute $w=c,\phi = \phie$ in the equation \eqref{fem5} and
%\bes
%\begin{split}
%&\gamma_1(x) =\eps_l(x),\gamma_2(x,w) =q_1(x,c),\gamma_3(x,w)=q_2(x,w),S(x,v,w)=0~\text{gives equation \eqref{fem1}},\\
%&\gamma_1(x) =0,\gamma_2(x,w) =q_3(x,c),\gamma_3(x,w)=q_4(x,w),S(x,v,w)=b(x)~\text{gives equation \eqref{fem2}},\\
%&\gamma_1(x) =0,\gamma_2(x,w) =0,\gamma_3(x,w)=1,S(x,v,w)=\frac{b(x)\jn}{\sigma_a}, \phi=\Phi_a~\text{gives equation \eqref{fem3}},\\
%&\gamma_1(x) =0,\gamma_2(x,w) =0,\gamma_3(x,w)=1,S(x,v,w)=\frac{b(x)\jn}{\sigma_c}, \phi=\Phi_c~ \text{gives equation \eqref{fem4}}.
%\end{split}
%\ees

The spatially discretised equations are obtained by using the approximation (\ref{fem0}) in (\ref{fem5}), multiplying by a test function $\psi_j(x)$, $j=0,...,N$, and integrating over the macroscopic domain $(0,1)$ to obtain
\be\label{fem7}
\begin{split}
	\sum_{i=0}^{i=N}\frac{dw_i}{dt}\int_{0}^{1}  \gamma_1(x) \psi_i \psi_j dx &= \left(\gamma_2(x,w)\frac{\partial w }{\partial x} +\gamma_3(x,w)\frac{\partial \phi }{\partial x} \right) \psi_j \bigg\rvert_{x=0}^{x=1} \\
	&-  
	\sum_{i=0}^N \left( w_i(t) \int_0^1 \gamma_2(x,w) \psi_i^{'}\psi^{'}_j dx +   \phi_i(t)\int_0^1\gamma_3(x,w) \psi_i^{'} \psi^{'}_j dx \right)\\
	&  + \int_0^1 S(x,v,w) \psi_j dx.
\end{split}
\ee 
The first-term on the right-hand side can be calculated using the appropriate boundary conditions, \eqref{fc21}-\eqref{fc24}. In general, the remaining integrals in \eqref{fem7} cannot be integrated exactly and further approximations are needed in order to progress. We adopt the approach given in \cite{Courtier18} and replace the functions $\gamma_1$, $\gamma_2$, $\gamma_3$ and $S$ appearing in the integrands by functions that are piecewise constant over each subinterval, $x\in(x_i, x_{i+1})$, and have a value equal to that of the function \eqref{fem0} at the midpoint of that interval. The first integral on the right-hand side of \eqref{fem7} is treated as follows 
\be\label{fem8}
\begin{split}
	\int_0^1 \gamma_1(x) \psi_i\psi_j dx &= \int_{x_{i-1}}^{x_i} \gamma_1(x) \psi_i\psi_j dx+ \int_{x_{i}}^{x_{i+1}} \gamma_1(x) \psi_i\psi_j dx\\
	&\approx  \gamma_1(x_{i-1/2}) \int_{x_{i-1}}^{x_i} \psi_i\psi_j dx+ \gamma_1(x_{i+1/2}) \int_{x_{i}}^{x_{i+1}} \psi_i\psi_j dx.
\end{split}
\ee
The first integral on the right-hand side in equation \eqref{fem7} can be approximated as follows
\be\label{fem9}
\begin{split}
	\int_0^1 \gamma_2(x,w) \psi_i^{'}\psi^{'}_j dx &= \int_{x_{i-1}}^{x_i} \gamma_2(x,w)\psi^{'}_i\psi_j^{'} dx+ \int_{x_{i}}^{x_{i+1}} \gamma_2(x,w) \psi_i^{'}\psi^{'}_j dx\\
	&\approx  \gamma_2\left(x_{i-1/2}, w\big\rvert_{x=x_{i-1/2}}\right) \int_{x_{i-1}}^{x_i} \psi_i^{'}\psi^{'}_j dx\\
	&+ \gamma_2\left(x_{i+1/2}, w\big\rvert_{x=x_{i+1/2}}\right) \int_{x_{i}}^{x_{i+1}} \psi_i^{'}\psi^{'}_j dx.
\end{split}
\ee
Treatment of second integral on right-hand side of \eqref{fem7} follows analogously but with $\gamma_2(x,w)$ replaced by $\gamma_3(x,w)$. Finally we approximate the final integral in \eqref{fem7} by writing
\be\label{fem10}
\begin{split}
	\int_0^1 S(x,v,w) \psi_j dx &\approx \frac{\Delta_{j-1/2}}{2} S\left(x_{j-1/2}, v\big\rvert_{x=x_{j-1/2}}, w\big\rvert_{x=x_{j-1/2}}\right)\\
	&+ \frac{\Delta_{j+1/2}}{2} S\left(x_{j+1/2}, v\big\rvert_{x=x_{j+1/2}}, w\big\rvert_{x=x_{j+1/2}} \right)
\end{split}
\ee
The errors incurred in using these approximations are second order (\ie their error decays proportional to the square of the grid spacing), just like the piecewise linear approximation for the dependent variables embedded in \eqref{fem0}. Hence the finite element discretisation retains its second order convergence rate despite the additional approximations. {The integrals on the right-hand side of equations \eqref{fem8}-\eqref{fem9} have integrands that depend solely upon the basis functions and their derivatives, and so can be computed exactly (for details see Appendix A). This observation leaves us in a position to write down the DAE system arising from the spatial discretisation of the macroscopic PDEs \eqref{fc1}-\eqref{fc4} of the Doyle-Fuller-Newman model. }

\subsection{Finite element implementation}
In order to write down the spatially discretised system of equations in a concise form we introduce three discrete operators: a difference operator $\mathcal{D}_i$, an operator for evaluation of dependent variables at a mid point $\mathcal{J}_i$ and a linear operator $\mathcal{L}_i$. These act on a column vector $\bf w$ with the entries 
\be\label{wi}
w_i=w|_{x=x_i}, \quad \text{for} \quad i=0,...,N
\ee
for a generic dependent variable $w$ they are defined as follows:
\be\label{D}
\begin{split}
	\frac{\partial w}{\partial x} \bigg\rvert_{x=x_{i+1/2}} &\approx \mathcal{D}_{i+1/2}({\bf w})=\frac{w_{i+1} -w_i}{\Delta_{i+1/2}}\\
	\label{J}
	w\big\rvert_{x=x_{i+1/2}}& \approx \mathcal{J}_{i+1/2}({\bf w}) = \frac{w_{i+1} + w_i}{2}\\
	\mathcal{L}_i({\bf w}) &= \frac{1}{6}\Delta_{i+1/2}w_{i+1} + \frac{1}{3}(\Delta_{i+1/2}+\Delta_{i-1/2})w_i+ \frac{1}{6}\Delta_{i-1/2}w_{i-1}.
\end{split}
\ee
Let $\bf x$  be a column vector of nodal points with the $i^\text{th}$ entry $x_i$ for $i=0,...,N$. {We seek to predict the electrolyte lithium concentration $c(x,t)$ and so, following (\ref{fem0}) and (\ref{wi}), we aim to find ${\bf c}(t)$ whose $i^{\text{th}}$ entry is $c_i=c(x_i,t)$ for $i=0,...,N$. The same is true for the electrolyte whose time-dependent values at the $N+1$ nodes $x_i$, $i=0,1,...,N$, are collated in the column vector ${\bf \Phi}(t)$. Similarly for the anode and cathode potentials whose time-dependent values at the $N_a+1$ and $N_c+1$ nodes respectively are collated in the column vectors ${\bf \Phi_a}(t)$ and ${\bf \Phi_c}(t)$. In addition, the values of the four quantities $q_k( x, c)$ (for $k=1,..,4$) at the $N+1$ nodes $x_i$, $i=0,1,...,N$ are stored in the vectors ${\bf q}^{(k)}(t)$.}

{We are now in a position to write down the spatially discretised equations arising from the macroscopic PDEs \eqref{fem1}-\eqref{fem4} and their boundary conditions (\ref{fc21})-(\ref{fc24}). We begin with the ODEs that govern the evolution of the lithium concentration in the electrolyte, and which are obtained from the spatial discretisation of \eqref{fem1} and  boundary conditions (\ref{fc21}b) and (\ref{fc24}b). These take the form}
\be\label{fem16}
\begin{split}
	& \Delta_{1/2}\left[\frac{1}{3}\frac{dc_0}{dt} + \frac{1}{6}\frac{dc_1}{dt} \right] = -\frac{1}{\eps_{1/2}} N_{1/2}\\
	\label{fem17}	
	& \mathcal{L}_i \left(\frac{d{\bf c}}{dt}\right)  = -\left[\frac{1}{\eps_{i+1/2}} N_{i+1/2}-\frac{1}{\eps_{i-1/2}} N_{i-1/2}\right],~\text{for} ~ i=1,...,N-1\\
	\label{fem18}
	& \Delta_{N-1/2}\left[\frac{1}{6}\frac{dc_{N-1}}{dt} + \frac{1}{3}\frac{dc_N}{dt} \right] =  \frac{1}{\eps_{N-1/2}}N_{N-1/2}.\\	
\end{split}	
\ee
{where $N_{i+1/2}$ is given by}
\be \label{fem11}
N_-|_{x=x_{i+1/2}} & \approx N_{i+1/2} = -\left[ \mathcal{J}_{i+1/2}( {\bf q}^{(1)}) \mathcal{D}_{i+1/2}({\bf c})   +   \mathcal{J}_{i+1/2}( {\bf q}^{(2)})\mathcal{D}_{i+1/2}({\bf \Phi}) \right]
\ee 
and {$\eps_{i+1/2} = \eps|_{x=x_{i+1/2}}$}. The algebraic equations for the  electrolyte potential $\Phi$, which result from the discretisation of equation \eqref{fem2} and the boundary conditions (\ref{fc21}c) and (\ref{fc24}c), are
\be\label{fem19}
\begin{split}
	0&= \phie_0\\
	\label{fem20}	
	0  &= -\left[  j_{i+1/2} - j_{i-1/2} \right] +{b_{i+1/2}} \frac{\Delta_{i+1/2}}{2} j^n_{i+1/2}  + { b_{i-1/2}}\frac{\Delta_{i-1/2}}{2} j^n_{i-1/2}, ~ \text{for} ~ i=1,...,N-1\\
	\label{fem21}
	0&= j_{N-1/2} +\frac{\Delta_{N-1/2}}{2} j^n_{N-1/2}, \\	
\end{split}	
\ee
where $j_{i+1/2}$ and $j^n_{i+1/2}$ are given by
\be
\label{12} j|_{x=x_{i+1/2}} \approx j_{i+1/2} = \left[ \mathcal{J}_{i+1/2}( {\bf q}^{(3)}) \mathcal{D}_{i+1/2}({\bf c})   +   \mathcal{J}_{i+1/2}( {\bf q}^{(4)})\mathcal{D}_{i+1/2}({\bf \Phi}) \right],\\
\jn\big\rvert_{x=x_{i+1/2}}  \approx j^n_{i+1/2} = \jn\left(\mathcal{J}_{i+1/2}({\bf c}), c_a|_{r=R_a(x_{i+1/2})}, c_c|_{r=R_c(x_{i+1/2})},\mathcal{J}_{i+1/2}({\bf \Phi}_a) , \mathcal{J}_{i+1/2}({\bf \Phi}_c)\right),
\ee
and $b_{i+1/2} = b|_{x=x_{i+1/2}}$}. The first equation in (\ref{fem19}) is required to set a reference value for the potential, and we select the value of zero at $x=L_1$ for convenience and without loss of generality. The algebraic equations for the  potential in anode $\Phi_a$, which result from discretisation of equation \eqref{fem3} and the boundary conditions (\ref{fc21}a) and \eqref{fc22}, are
\be\label{fem22}
\begin{split}
	0&=-\frac{I}{A}+j^a_{1/2} - \frac{\Delta_{1/2}}{2} b_{1/2} j^n_{1/2}\\
	\label{fem23}	
	0  &=- \left[  j^a_{i+1/2} - j^a_{i-1/2} \right] - \frac{\Delta_{i+1/2}}{2} b_{i+1/2} j^n_{i+1/2}  - \frac{\Delta_{i-1/2}}{2}b_{i-1/2} j^n_{i-1/2}, ~ \text{for} ~ i=1,...,N-1\\
	\label{fem24}
	0&= -j^a_{N-1/2} -\frac{\Delta_{N-1/2}}{2} b_{N-1/2}j^n_{N-1/2}, \\	
\end{split}	
\ee
where 
\be
j_a|_{x=x_{i+1/2}} & \approx j_{i+1/2}^a = -\sigma_a \mathcal{D}_{i+1/2}({\bf \Phi}_a).
\ee
The algebraic equations for the  potential in cathode $\Phi_c$, which result from discretisation of equation \eqref{fem3} and the boundary conditions (\ref{fc23}) and (\ref{fc24}a), are
\be\label{fem25}
\begin{split}
	0&=-j^c_{1/2} - \frac{\Delta_{1/2}}{2} b_{1/2} j^n_{1/2}\\
	\label{fem26}	
	0  &= -\left[  j^c_{i+1/2} - j^c_{i-1/2} \right] - {b_{i+1/2}}\frac{\Delta_{i+1/2}}{2} j^n_{i+1/2}  -  {b_{i+1/2}}\frac{\Delta_{i-1/2}}{2} j^n_{i-1/2}, \quad \text{for} \quad i=1,...,N-1\\
	\label{fem27}
	0&= -\frac{I}{A}+j^c_{N-1/2} - b_{N-1/2}\frac{\Delta_{N-1/2}}{2} j^n_{N-1/2}. \\	
\end{split}	
\ee
where
\be \label{stink}
j_c|_{x=x_{i+1/2}} & \approx j_{i+1/2}^c = -\sigma_c \mathcal{D}_{i+1/2}({\bf \Phi}_c).
\ee
Equations (\ref{fem18})-(\ref{stink}) comprise the discretised macroscopic equations that are implemented in DandeLiion.

\subsection{Assembly of the Differential Algebraic Equations}
Here we briefly describe how the system of DAEs, which are solved by DandeLiion, are assembled from the spatially discretisation of the DFN model. As stated earlier the microscopic equations \ma{\eqref{fc30}-\eqref{fc31}} are discretised by application of  Zeng \etal's \cite{Zang13} control volume (CV) method, which like the FEM discretisation of the macroscopic equations, exhibits perfect lithium conservation and also provides directly evaluates the lithium-ion concentration on the electrode particle surfaces, which is important from the point of view of accurately approximating the Butler-Volmer equations.

{Henceforth we refer to the combined finite element and control volume spatial discretisation as the FE+CV scheme.}  The total number of grid points in the macroscopic dimension, $x$ is $N= N_a+N_s+N_c$, where $N_a$, $N_s$, $N_c$ are the grid points in the anode, separator, and cathode respectively. At each of the $N_a+N_c$ stations in $x$ which belong to the anode or cathode we consider {a representative spherical electrode particle which is discretised using $M$ grid points in the radial coordinate $r$}. We denote $r_j=jh_r$, where  $h_r = 1/(M-1)$ for $j=1,...,M$. In total we have $(N_a +N_c)\times M$ different stations in $r$ and at these locations we denote the value of lithium concentration in anode and cathode by $c^a_{i,j}$ and  $c^c_{i,j}$ respectively. The index $i$ indicates the representative particle’s position in $x$ whereas $j$ labels the radial position within that particle. In total we have $2N$ functions to be determined for concentration and potential in the electrolyte, $N_a$ and $N_c$ unknowns for the potential in anode and cathode respectively, and $(N_a+N_c)\times M$ unknowns for the concentration in anode and cathode.

{The $2N+(N_a+N_c)\times (M+1)$ unknown functions of time are assembled into one large column vector ${\bf u}(t)$ as follows
\be\label{fem28} \begin{split}
{\bf u}(t)= \left[c_0,...c_N,\phie_0,...,{\phie_N,\phia^0},...\phia^{N_a},\phic^0,...,\phic^{N_c} ,c_a^0,...c_a^{N_a}, c_c^0,...c_c^{N_c}\right]^T \\ = \left[{\bf c}(t)^{T}  {\bf \Phi}(t)^{T}  {\bf \Phi}_a(t)^{T}    {\bf \Phi}_c(t)^{T}   {\bf c_a}(t)^{T},{\bf c_c}(t)^{T}\right]^T \end{split}
\ee
where the superscript $T$ denotes a transpose. This allows the system of DAEs to be written in the concise form 
\be\label{30}
{\bf M} \frac{d{\bf u}}{dt}= {\bf f(u)},\quad \text{with}\quad {\bf u}\big\rvert_{t=0}={\bf u}_0.
\ee
Here the mass matrix ${\bf M}$ is a $ (2N+(N_a+N_c)\times (M+1))\times (( 2N+(N_a+N_c)\times (M+1))$ tridiagonal matrix whose entries are coefficients of the time derivative terms in the equations \eqref{fem17}--\eqref{fem27}, and control volume descretisation from \cite{Zang13}. The vector function ${\bf f(u)}$ is nonlinear, and has length $2N+(N_a+N_c)\times (M+1)$. Its entries are the right-hand sides of equations \eqref{fem17}--\eqref{fem27} and the equations arising from the control volume descretisation. The DAE system (\ref{30}) is integrated forward in time using DandeLiion's {in house} DAE solver.} 

\section{Verification}
{In this section, we demonstrate the second order convergence of our FE+CV method by benchmarking against an alternative spatial discretisation applied to the DFN model. We select a standard finite volume method, see \cite{LeVequebook}, to compare against and we apply this spatial discretisation to both the macroscopic and microscopic components of the model, \ie \eqref{fc1}-\eqref{fc33}. As such, we will henceforth refer to this approach as the FV+FV method which is expected to, and indeed does, exhibit first order convergence.} Since the FE+CV method is comprised of a combination of two different methods for spatial discretisation (finite elements and control volumes) we will validate the overall second order convergence rate in two steps. First we demonstrate that the application of the CV method to a nonlinear spherical diffusion equation exhibits second order converge as the number of grid points $M$ is increased. Then, we verify the second order convergence rate for the FE discretisation by refining the number of grid points $N$, {in the macroscopic dimension $x$}, whilst taking $M$ {the number of grid points in the microscopic dimension $r$ to be large enough such} that the numerical errors arising from the {discretisation of the} microscopic equations are negligible. An analogous two-stage strategy is used for the FV+FV method. Throughout all our spatial convergence testing we set the error tolerances on the DAE integrator to be sufficiently stringent that time integration errors can also be assumed to be negligible.

Our benchmarking protocol will be based on a cell parameterised with the data in Ecker \etal \cite{Ecker2015a,Ecker2015b} for a single full discharge cycle at 4C. Due to the lack of an exact solution a reference solution, computed on a very refined grid, is used to assess the errors. For some scalar quantity $w$ (which could be $c, \phie, \phia, \phic,c_a,c_c$ evaluated at fixed spatial and temporal values, we can define the numerical error of a simulation as
\be\label{errorN}
\mathcal{E}(w,N,N_{\text{ref}})= | w^{(N)}- w^{(N_{ref})} |,
\ee
where $| \cdot |$ is the absolute value operator and $w^{(N_{ref})}$ is the {approximation to the exact solution found by using a highly refined grid}.

%\red{(IT IS NOT AT ALL CLEAR WHAT YOU ARE DOING. HOW DO YOU TAKE THE NORM. DO YOU TAKE IT IN SPACE OR IN TIME, OR BOTH. IF ITS NOT CLEAR YOU NEED TO DEFINE THE NORM HERE)}

% At fixed time $t=T_{t}$ we call $\mathcal{E}^1(w,N,N_{\text{ref}})$ absolute error note that here $w$ is a scalar quality. In this section, we chose the red colour to show the error plots generated by using FV and FV+FV methods and blue colour for CV and FE+CV methods. In Table \ref{order_solid}-\ref{order_voltage}, we show the numerical order of convergence which we computed by using linear least square method on the data points from Figure \ref{errors_plots}. }

{We first investigate the dependence of the numerical convergence on $M$, the number of grid points used to discretize the particle diffusion equations \eqref{fc30}-\eqref{fc33}, for a fixed value of $j_n$}. We compute a good approximation to the exact solution solution by taking a large value, in this instance $M_{\text{ref}}=2561$. In Figure \ref{errors_plots}, the top left plot shows the {logarithm of} the absolute errors {$\mathcal{E}(c_a({r=R_a/2,t=t_f}),M,M_{\text{ref}})$} and {$\mathcal{E}(c_c({r=R_c/2,t=t_f}),M,M_{\text{ref}})$}, for $t_f=800$~s {plotted versus $\log(M)$} for the concentration in anode and cathode using FV and CV methods. We emphasize that in this test $j_n$ is taken to be constant and as such there is no need to evaluate $c_a$ or $c_c$ at a specified $x$ (they too are independent of $x$). As expected {the straight line fit to the CV method  has a gradient of $\approx -2$ corresponding to second-order accuracy of the scheme while the straight line fit to FV method has a gradient of $\approx -1$ corresponding to first-order accuracy of the scheme. More details of the calculation of the numerical order of convergence are given in Table \ref{order_solid}, which shows the value of minus the gradient of the least square straight line fits to the data, in \ref{errors_plots}, which corresponds to the order of convergence.}

%\red{(AGAIN IT IS NOT CLEAR WHAT YOU ARE DOING. DO YOU CALCULATE THE SOLUTION ONLY FOR A SINGLE PARTICLE? DO YOU ASSESS THE NORM AT A GIVEN TIME?)}

{Next we investigate the dependence of the numerical convergence on $N$, the number of grid points used to discretize the macroscopic DFN equations \eqref{fc1}-\eqref{fc2}.} We compute a reference solution, corresponding to a good approximation to the exact solution, by using a large number of grid points, in this case $N_{\text{ref}}= 1723$ for FE+CV method and $N=N_{\text{ref}}=2560$ for FV+FV method. {Throughout the tests to assess converge in $N$ we fix the number of grid points $M=640$ which is sufficiently large that errors stemming from the solution to the diffusion equations describing transport in the electrode particles are negligible}. 

In Figure \ref{errors_plots} the top right plot shows {the logarithms of the absolute errors for electrolyte concentration $c$, electrolyte potential $\Phi$ and anode potential ${\Phi_a}$ at the midpoint of the anode plotted against $\log(N)$ using the FE+CV method and using the FV+FV method}. {The same least square fitting procedure is used as above to assess the numerical order of convergence from the variation in error with radial grid spacing and the results are displayed in Figure \ref{errors_plots}. In particular it shown that FE+CV method is second order while the FV+FV is first order.}

{Finally, in the lower panel of Figure \ref{errors_plots}, we show the error in the output voltage} which is a function of time and can therefore be assessed using an error defined as
\be
\mathcal{E}^p(V(t),N,N_{\text{ref}}) = \big\| V(t)^{(N)} - V(t)^{(N_{ref})} \big\| 
\ee
and $p=1,2,\infty$. As expected the FE+CV shows second-order convergence, corresponding to a straight line with gradient -2 in the log($\varepsilon^p$) and FV+FV shows only first order convergence.

{
\begin{figure} \centering
	\includegraphics[width=0.44\textwidth]{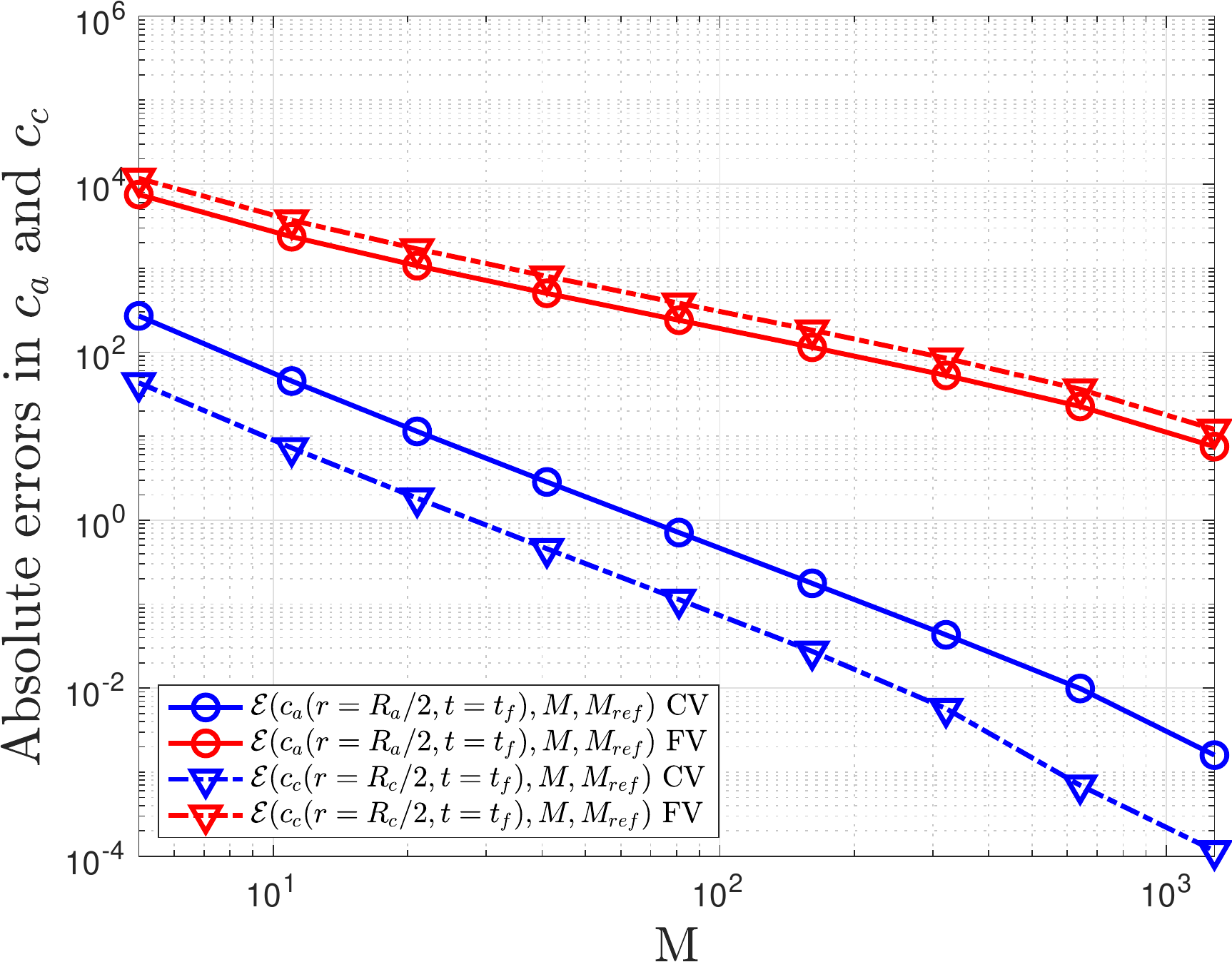}\hspace*{1.cm}
	\includegraphics[width=0.44\textwidth]{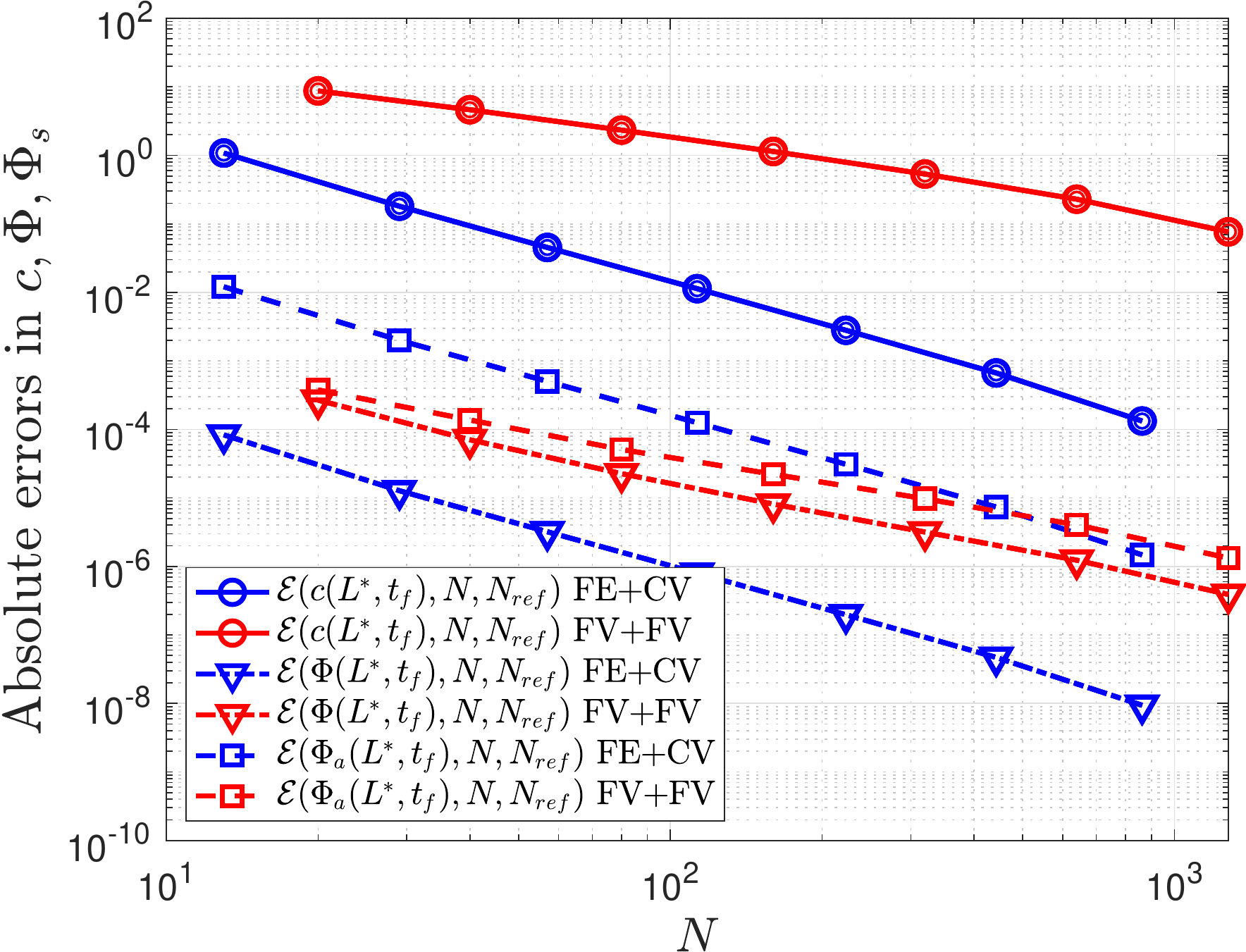}\hspace*{1.5cm}\\
	\includegraphics[width=0.44\textwidth]{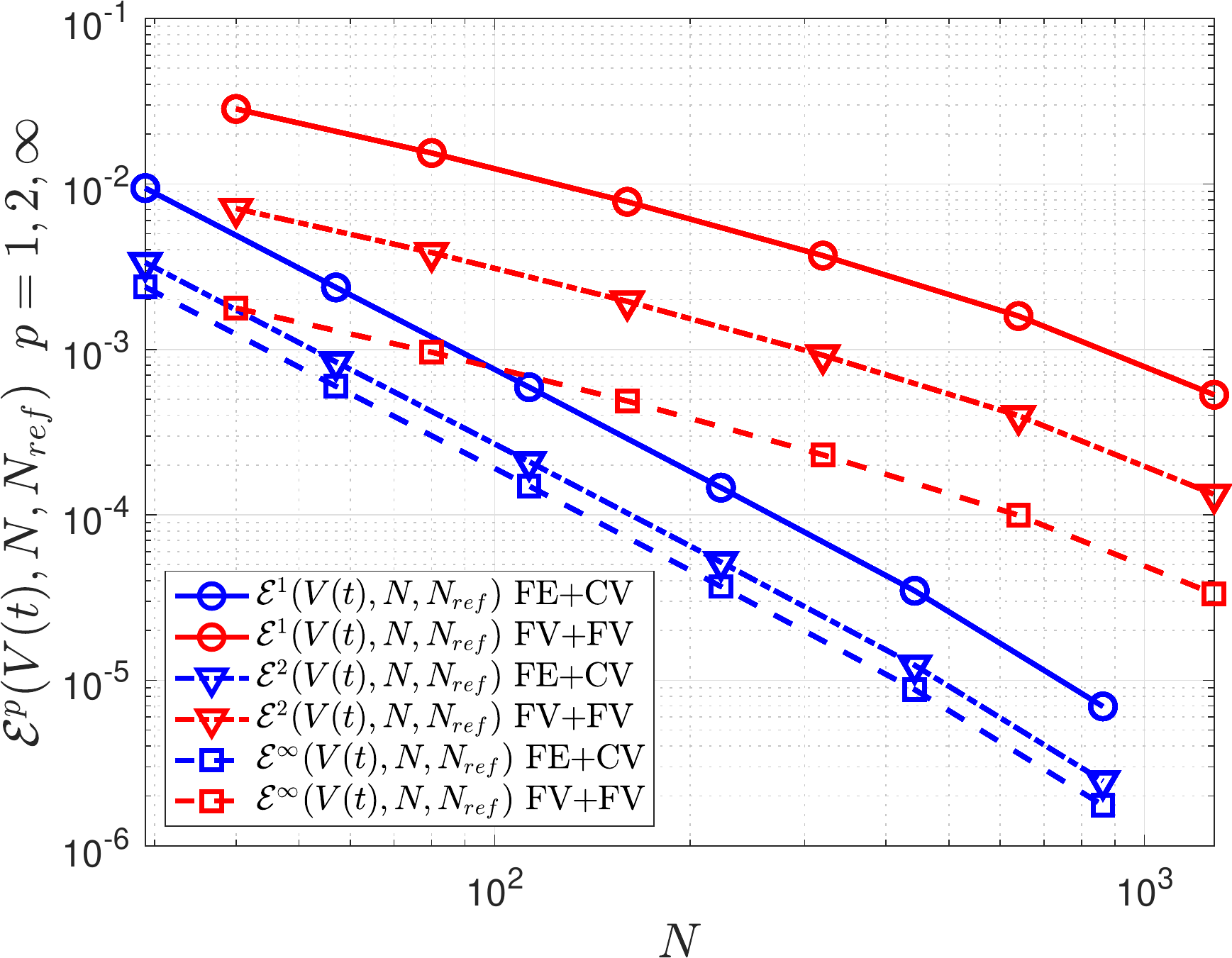}\\
	\caption{{The top left panel is a plot of the absolute errors $\mathcal{E}(c_a(r=R_a/2, t=t_f),M,M_{\text{ref}})$ and $\mathcal{E}(c_c(r=R_c/2, t=t_f),M,M_{\text{ref}})$ for concentration in anode and cathode with $t_f=800$~s. The top right panel is a plot of the absolute errors $\mathcal{E}(c(x=L^*,t=t_f),N,N_{\text{ref}}), \mathcal{E}(\Phi(x=L^*, t=t_f),N,N_{\text{ref}})$ and $\mathcal{E}({\Phi_a}(x=L^*, t=t_f),N,N_{\text{ref}})$ for concentration in electrolyte, potential in electrolyte and potential in solid $t_f=800$~s and $L^*= {(L_1+L_2)/2}$. The bottom panel is a plot of the errors $\mathcal{E}^p(V(t),N,N_{\text{ref}})$ where $p=1,2,\infty$  for the cell voltage.}}
	\label{errors_plots}
\end{figure}
\begin{table}[H]
	\centering
	\begin{tabular}{|c|c|c|}
		\hline \textbf{Method} &\textbf{Order for $c_a(r=R_a/2,t=t_f)$} & \textbf{Order for  $c_c(r=R_c/2,t=t_f)$}\\
		\hline
		FV & 1.19 &   1.19   \\
		\hline
		CV & 2.12 &   2.27  \\
		\hline
	\end{tabular}
	\caption{Numerical orders of convergence for concentration in anode $c_a(r=R_a/2,t=t_f)$ and concentration in cathode $c_c(r=R_c/2,t=t_f)$ by using FV and CV methods with $t_f=800$~s. The orders are arrived at by fitting a straight line (using least squares) to the data in Figure \ref{errors_plots}.}
	\label{order_solid}
\end{table}
\begin{table}[H]
	\centering
	\begin{tabular}{|c|c|c|c|}
		\hline \textbf{Methods} & \textbf{Order for} & \textbf{Order for}& \textbf{Order for} \\
		& $c(x=L^*,t=t_f)$ & $\Phi(x=L^*,t=t_f)$ & ${\Phi_a}(x=L^*,t=t_f)$ \\
		\hline
		FV+FV &1.05 &   1.53 &  1.30    \\
		\hline
		FE+CV&2.08 &   2.10 &  2.08  
		\\
		\hline
	\end{tabular}
	\caption{Numerical orders of convergence for concentration in electrolyte for $c(x=L^*,t=t_f)$, potential in electrolyte  $\Phi(x=L^*,t=t_f)$ and potential in solid $\Phi_s(x=L^*,t=t_f)$ by using FV+FV and FE+CV methods with $t_f=800$~s and $L^*= {(L_1+L_2)/2}$. The orders are arrived at by fitting a straight line (using least squares) to the data in Figure \ref{errors_plots}.} 
	\label{order_c_phi_phis}
\end{table}
\begin{table}[H]
	\centering
	\begin{tabular}{|c|c|c|c|}
		\hline 
		\textbf{Methods} &\textbf{Order for $V(t)$, $L^1$} & \textbf{Order for $V(t)$, $L^2$} & \textbf{Order for $V(t)$, $L^\infty$}  \\
		\hline
		FV+FV   & 1.04 &   1.04 &  1.04\\ 
		\hline
		FE+CV& 2.09 &   2.09 &  2.08 \\
		\hline
	\end{tabular}
	\caption{Numerical orders of convergence for voltage $V(t)$ in $L^1$, $L^2$ and $L^\infty$ norm by using FV+FV and FE+CV methods. The orders are arrived at by fitting a straight line (using least squares) to the data in Figure \ref{errors_plots}.}
	\label{order_voltage}
\end{table}

\section{Illustrative examples}
\label{examples}

To demonstrate the practical utility of DandeLiion we show a single discharge cycle based on Graphite-Silicon/$\text{Li}\text{Ni}_{1-x-y}\text{Mn}_x\text{Co}_{y}\text{O}_2$ LG~M50 battery cell chemistry \cite{Ferran} and a simulation of a charge/discharge current profile applied to the cell at different (dis)charge rates.

The DFN model implemented in DandeLiion was fully parametrised according to \cite{Ferran}. All the parameters, including functions (\eg open circuit voltages, diffusivity and conductivity in the electrolyte, see Figure~\ref{fig:preview}) were filled directly in the web forms provided by the simulation engine on the DandeLiion website \cite{sim_form}. The computational grid can be defined by the user as well, and for the purposes of this demonstration we set up 50 grid points in the electrolyte in each electrode, 30 points across the separator, and 100 nodes in each solid particle. The authors in \cite{Ferran} test their parametrisation using 0.5C, 1C, and 1.5C constant discharge currents followed by a relaxation period. As a first example, we simulate a full 1C discharge with two-hour relaxation. For the chosen discretisation the total number of DAEs to be solved reaches 10$^4$, but the compute time remains very managebale at around 1 second. In addition, when running this on the server there is a fixed (independent of simulation size/complexity) overhead of around 7-10 seconds which is associated with setting up the simulation in the cloud, checking the user-defined parameterisation, code compilation, saving the data, creating a zip archive and generating a permanent webpage displaying the results.

%\red{(THIS SEEMS A LONG TIME, ALMOST AS LONG AS COMSOL OR DYMOLA FOR A RELATIVELY SIMPLE TASK. NEED TO EXPLAIN WHY THIS IS OR PERHAPS TAKE A COARSER GRID?)} \bl{\it{On SOTON server (which has more cores) it works faster, the simultaion time is 20 sec now. We will move our online solver to SOTON server in December. Reducing the grid size does not help much, because of overheads explaining below}} which includes the setting up of the job, checking of the user-defined parametrisation, compiling the simulation, perform the calculation, generating the webpages with the plots, and producing a \bl{permanent} link where the user can download the results. \bl{It should be noted that the compute time itself is approximately one second. The remainder of the time is related to setting up the simulation in the cloud and to the code compilation using the user-defined parametrisation. Regardless to the simulation complexity, these overheads are the same (about 20 seconds).}

%Immediately after the submission of the simulation on the server, the code performs basic parametrisation check. If it successful, the server shows plots of the functions defined by the user (Figure~\ref{fig:preview}) and the current status of the job.

\begin{figure} \centering
\includegraphics[width=0.495\textwidth]{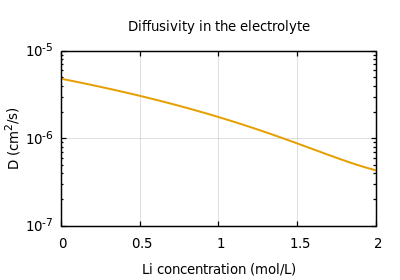}
\includegraphics[width=0.495\textwidth]{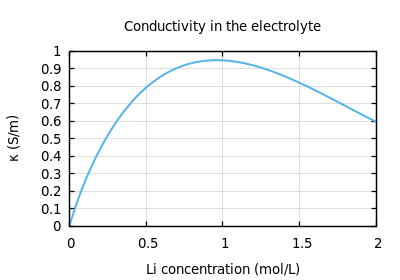}\\
\medskip
\includegraphics[width=0.495\textwidth]{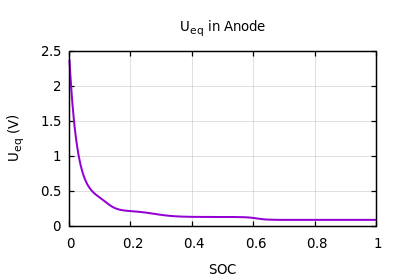}
\includegraphics[width=0.495\textwidth]{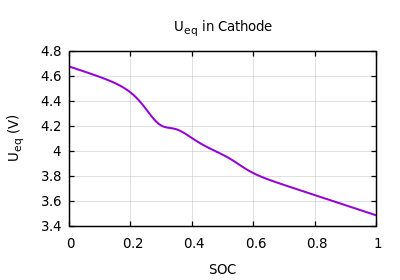}\\
\caption{{Ionic diffusivity (top left) and conductivity (top right) of the electrolyte, open circuit voltages of graphite-silicon anode (bottom left) and $\text{Li}\text{Ni}_{1-x-y}\text{Mn}_x\text{Co}_{y}\text{O}_2$ cathode (bottom right). Experimental data from from \cite{Ferran}.}}
\label{fig:preview}
\end{figure}

After the job is complete DandeLiion users can see a set of preliminary plots (the output of the simulation described here is shown in Figure~\ref{fig:multiplot}). These plots show the total voltage and user-defined current against time, as well as the Li ion concentration in the electrolyte and within two representative particles; one in the anode and another in the cathode, as well as the potential distribution in the electrolyte. Below these plots, a link is provided to download the raw data files for plotting using any other software of choice, \eg Microsoft Excel, MATLAB, etc. 

This simulation was used to further validate DandeLiion. The voltage is compared with both experiment and simulation results from \cite{Ferran} and Figure~\ref{fig:volt_1C} shows that good agreement is obtained.

\begin{figure} \centering
\includegraphics[width=0.95\textwidth]{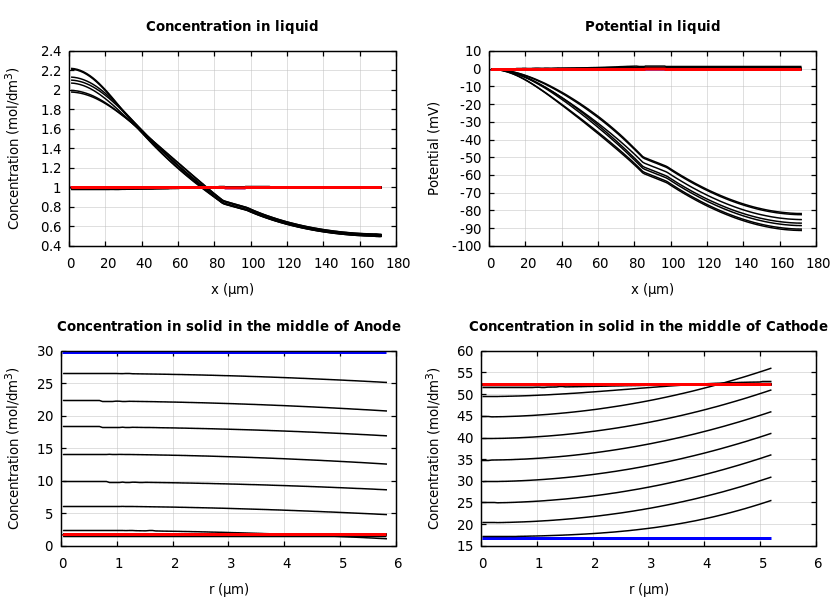}\\
\caption{{Concentration of Li (top left) and potential distribution (top right) in the electrolyte across the battery cell width, and concentrations of Li in two representative particles in anode (bottom left) and cathode (bottom right). The blue line shows the initial state of the battery, black thin lines correspond to the snapshots at 500~s intervals, and the red line is the final state.}}
\label{fig:multiplot}
\end{figure}

\begin{figure} \centering
\includegraphics[width=0.7\textwidth]{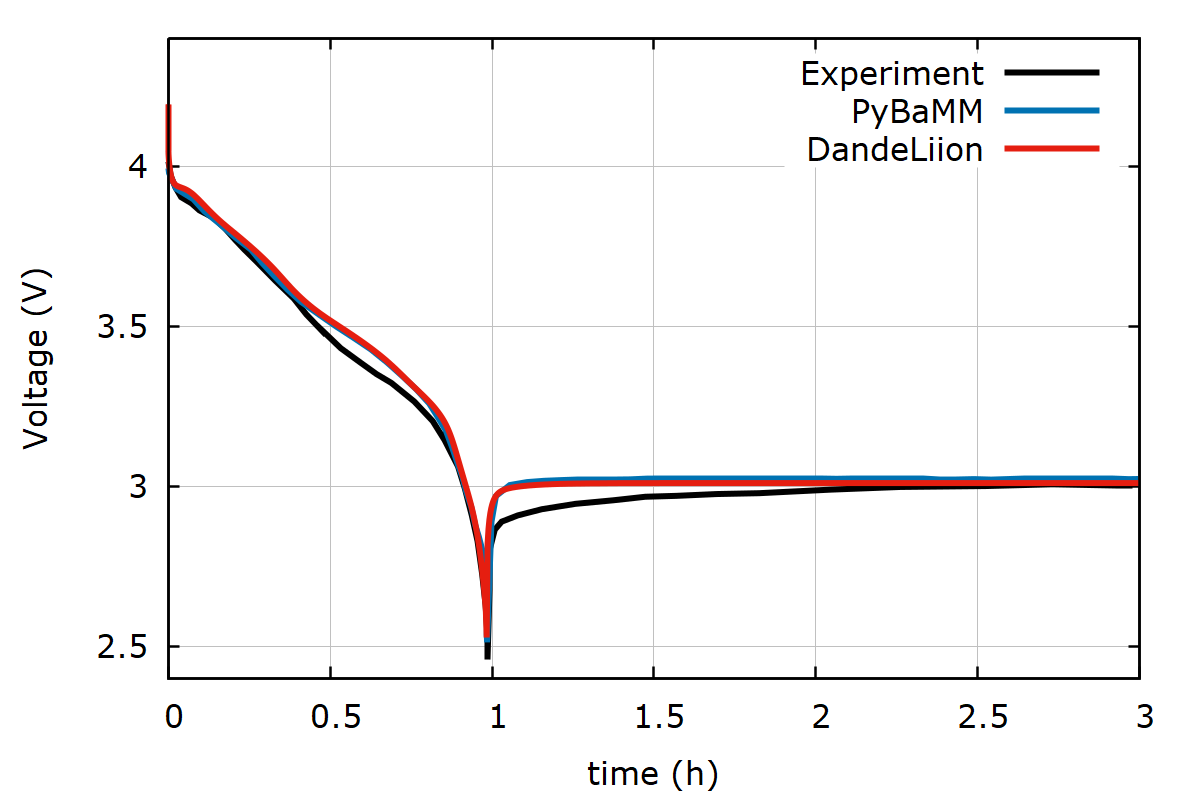}\\
\caption{{Cell voltage profiles computed from DandeLiion in comparison with the experimental and simulation data from \cite{Ferran}.}}
\label{fig:volt_1C}
\end{figure}

{After the simulation is complete, the user may change any of the parameters and resubmit the simulation (this can be done even {\it during} the simulation, a new instance of the simulation will be created and sent into the queue). There is no need to complete the parametrisation form from scratch; all parameters are stored on the server and can be re-used by clicking on the `Review all parameters \& Resubmit the simulation' button. The server will create a permanent link for each parametrisation so that it can be bookmarked for future use.}

{DandeLiion allows the user to define different particle sizes in each electrode thereby allowing simulation of so-called graded electrodes which might have larger particles adjacent to the separator than those near the current collector, or vice versa. As a demonstration of this functionality, we take the parameter set in \cite{Ferran} and increase the particle size in anode near the separator by a factor of three so that those particles in $L_1<x<(0.1L_1+0.9L_2)$ are of size $R_a$ and those in $(0.1L_1+0.9L_2)<x<L_2$ are of size $3R_a$. The increased size of particle was accommodated in the electrode by decreasing the number of particles, as well as the particle surface area (per unit volume) $b(x)$, in $(0.1L_1+0.9L_2)<x<L_2$ by the same factor of three. The inclusion of the graded electrode functionality is motivated by the clear variation in particle sizes seen in microscopy data of real electrodes, see \cite{Ferran,Liu} for examples. The importance of capturing these variations is spoken to by the quality of the agreement between DandeLiion and experiment \cite{FerranData} shown in Figure~\ref{fig:volt}. We emphasize the improvement in fit between Figures~\ref{fig:volt} and \ref{fig:volt_1C} is due to the variation in particle sizes that is accounted for in the former, but not the latter. 

Both simulation examples including the parametrisation and corresponding current profiles are available on the DandeLiion website \cite{sim_form}.

\begin{figure} \centering
\includegraphics[width=0.6\textwidth]{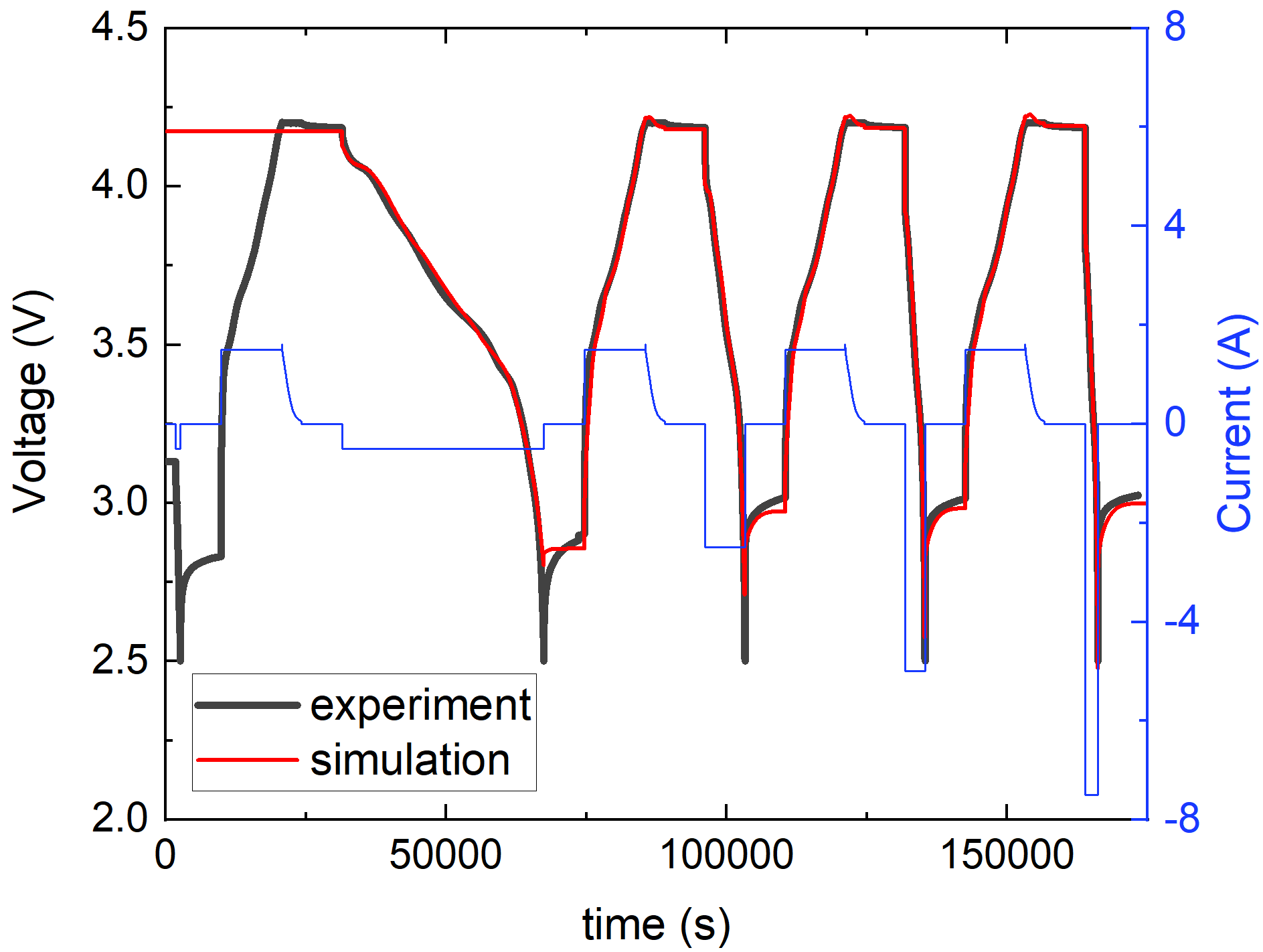}\\
\caption{{Cell voltage and the current vs time comparing DandeLiion simulations and experiment taken from \cite{FerranData}. The current varies between (dis)charge rates ranging between $\pm$1.5C.}}
\label{fig:volt}
\end{figure}

\section{Conclusions}
\label{conclusions}
This work describes the release of novel software that is able to solve the most ubiquitous electrochemical cell-scale LIB model, namely the DFN model, extremely quickly. DandeLiion is a cloud-based service, accessible via {\tt dandeliion.com}, where users can submit their jobs via an easy-to-use web interface and can collect results both in the browser and in-full by downloading raw output. It comes equipped with comprehensive documentation, a set of video tutorials aimed at new users and a library of chemistries to construct common cell architectures. A set of pre-defined simulations are available on the website that can be adapted to suit user's specific needs. In the future we aim to expand upon the existing material library, and add additional physics including thermal coupling across multiple cells. 
%We are open to suggestions on additional functionality that could be added to serve the community.

DandeLiion has the capability of making rapid predictions of LIB (dis)charge behaviour and arms both academics and industrialists with the means of solving a model which has been demonstrated to accurately predict device behaviour across a range of operating protocols and a variety of device designs \cite{RicRev,Jok}. {The ability to solve this model much more rapidly than previously opens the door to being able to investigate multi-dimensional thermally coupled problems in composite cells (\eg pouch cells and cylindrical cells), battery modules and even in entire battery packs, using a realistic electrochemical representation of the cell (rather than relatively crude equivalent circuit models). It will also enable modern optimisation techniques to be applied to electrochemical models of the cell and used to design optimal cell structures and furthermore it opens the way to using parameter estimation techniques to  deduce cell properties from real cell data.} It also facilitates finding solutions in computationally intensive settings, such as a realistic drive cycle. 

{DandeLiion's functionality expedites the development of new device designs by allowing users to explore the effects of alterations to battery designs in-silico, lowering the monetary and temporal costs associated with development via physical prototyping. It therefore paves the way for significant improvements in LIB performance, lifetime and safety, especially in the context of their use in EVs and other high-power applications. Ultimately this significant advance in LIB simulation software is expected to  lead to substantial benefits to industry and increase the impetus for the creation of new products and procedures.}

\section{Conflict of Interest}
We wish to confirm that there are no known conflicts of interest associated with this publication and there has been no significant financial support for this work that could have influenced its outcome.

\section*{Acknowledgements}
\label{acknowledgements}
The work of all the authors was supported by the Faraday Institution Multi-Scale Modelling (MSM) project (grant number EP/S003053/1). The authors would like to thank Debora Corbin for suggesting the name of the software as well as Ferran Brosa Planella and Emma Kendrick for providing the experimental data (on the LG~M50 battery) used here for validation.

%\red{Please add the reference to the software repository if DOI for software is available.}

\section{Appendix A.  Integrals required for the finite element descretisation}\label{sec:AppendixA}
In section \ref{fem}. when using the finite element method to descretise the governing system of equations  \eqref{fc1}-\eqref{fc4} in space the following results are needed:
\be\label{A1}
\int_{0}^{1}\psi_j dx = \left\{
\begin{array}{ll}
	\frac{1}{2} (\Delta_{j+1/2}+\Delta_{j-1/2})& ~\text{if}~ j=1,..,N-1 \\
	\frac{1}{2} \Delta_{1/2}& ~\text{if}~ j=0 \\
	\frac{1}{2} \Delta_{N-1/2}& ~\text{if}~j=N\\
	0 & \text{otherwise}\\
\end{array} 
\right. 
\ee
\be\label{A2}
\int_{0}^{1}\psi_i \psi_j dx = \left\{
\begin{array}{ll}
\frac{1}{3} (\Delta_{j+1/2}+\Delta_{j-1/2})& ~\text{if}~ i=j ~\text{and}~ j=1,..,N-1 \\
\frac{1}{3} \Delta_{1/2}& ~\text{if}~ i=j ~\text{and}~ j=0 \\
\frac{1}{3} \Delta_{N-1/2}& ~\text{if}~ i=j ~\text{and}~ j=N\\
\frac{1}{6} \Delta_{j+1/2}& ~\text{if}~ i=j+1~\text{and}~ j=0 ,...,N-1\\
\frac{1}{6} \Delta_{j-1/2}& ~\text{if}~ i=j-1~\text{and}~ j=1,...,N\\
0 & \text{otherwise}\\
\end{array} 
\right. 
\ee
\be\label{A3}
\int_{0}^{1}\psi_i^{'} \psi^{'}_j dx = \left\{
\begin{array}{ll}
\frac{1}{\Delta_{j+1/2}}+\frac{1}{\Delta_{j-1/2}}& ~\text{if}~ i=j ~\text{and}~ j=1,..,N-1 \\
\frac{1}{\Delta_{1/2}}& ~\text{if}~ i=j ~\text{and}~ j=0 \\
\frac{1}{\Delta_{N-1/2}}& ~\text{if}~ i=j ~\text{and}~ j=N\\
\frac{-1}{\Delta_{j+1/2}}& ~\text{if}~ i=j+1~\text{and}~ j=0 ,...,N-1\\
\frac{-1}{\Delta_{j-1/2}}& ~\text{if}~ i=j-1~\text{and}~ j=1,...,N\\
	0 & \text{otherwise}\\
\end{array} 
\right.,	 
\ee
%\be\label{A4}
%\int_{0}^{1}\psi^{'}_i \psi_j dx = \left\{
%\begin{array}{ll}
%	%\frac{1}{2} & ~\text{if}~ i=j ~\text{and}~ j=1,..,N-1 \\
%    -\frac{1}{2} & ~\text{if}~ i=j ~\text{and}~ j=0 \\
%	\frac{1}{2}& ~\text{if}~ i=j ~\text{and}~ j=N\\
%	\frac{1}{2} & ~\text{if}~ i=j+1~\text{and}~ j=0 ,...,N-1\\
%   -\frac{1}{2} & ~\text{if}~ i=j-1~\text{and}~ j=1,...,N\\
%	0 & \text{otherwise}\\
%\end{array} 
%\right. ,
%\ee
where $\psi_i(x)$ is a basis function as defined in \eqref{fem0}, a prime denotes a derivatives with respect to $x$, and the indices $0\leq i,j \leq N$.
\end{document}